*Review*

# Twenty-Year Review of Outdoor Air Quality in Utah, USA

Callum E. Flowerday [1], Ryan Thalman [2] and Jaron C. Hansen [2,*]

1 Department of Chemistry and Biochemistry, Brigham Young University, Provo, UT 84602, USA
2 Department of Chemistry, Snow College, Richfield, UT 84701, USA
* Correspondence: jhansen@chem.byu.edu; Tel.: +1 801-422-4066

**Abstract:** Air quality is a prevalent concern due to its imposing health risks. The state of Utah, USA, at times over the last 20 years has suffered from some of the worst air quality in the nation. The propensity for the state of Utah to experience elevated concentrations of particulate matter and ozone can in part be attributed to its unique geography that features dry, mountainous topography. Valleys in Utah create ideal environments for extended cold-pool events. In this review, we summarize the research executed in Utah over the past 20 years (2002–2022) by dividing the state into six regions: Utah Valley, Summit County, Southern Utah (regions south of Utah Valley), Cache Valley, Uinta Basin, and Salt Lake Valley. We review the published literature chronologically and provide a summary of each region identifying areas where additional research is warranted. We found that the research effort is weighted towards Uinta Basin and Salt Lake Valley, with the other regions in Utah only adding up to 20% of the research effort. We identified a need for more source apportionment studies, speciated volatile organic compound (VOC) studies, and ozone isopleths. Where ozone isopleths are not able to be created, measurement of glyoxal and formaldehyde concentrations could serve as surrogates for more expensive studies to inform ozone mitigation policies.

**Keywords:** tropospheric ozone; regional contribution to ozone; local contribution to ozone; ozone isopleths; atmospheric oxidants

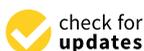

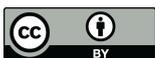

## 1. Introduction

Air quality is a much-discussed and researched issue in the world because of its effects on health and the climate [1,2]. Poor air quality is estimated to be responsible for 7 million premature deaths globally every year. As such, researchers have invested substantial efforts into both identifying and quantifying atmospheric pollutants. Utah, USA, has had problems in the past with air quality, including elevated $O_3$, and $PM_{2.5}$ concentrations, some of which are exacerbated due to the geographical profile of the region [3]. The geography in northern and central Utah can produce prolonged cold pool events that promote the formation of secondary particulate matter [4–6]. Over the last 20 years (2002–2022) the population in the state of Utah has increased by 44% from 2,324,815 (2002) to 3,337,975 (2022) and with the increased population there has been an increase in energy consumption, vehicle emissions, and other anthropogenic emissions. It is generally expected that increases in gross domestic product (GDP), vehicle miles traveled (VMT), population, and energy consumption result in increased emissions. Figure 1 shows the areas of growth in Utah between 2000–2022 normalized to their values in the year 2000. It shows that despite a growth in GDP, VMT, population, energy consumption, and $CO_2$ emissions, aggregate emissions, calculated by summing the concentrations of CO, PM, $NO_2$, $O_3$ and $SO_2$, that air quality has improved in the state since the year 2000.

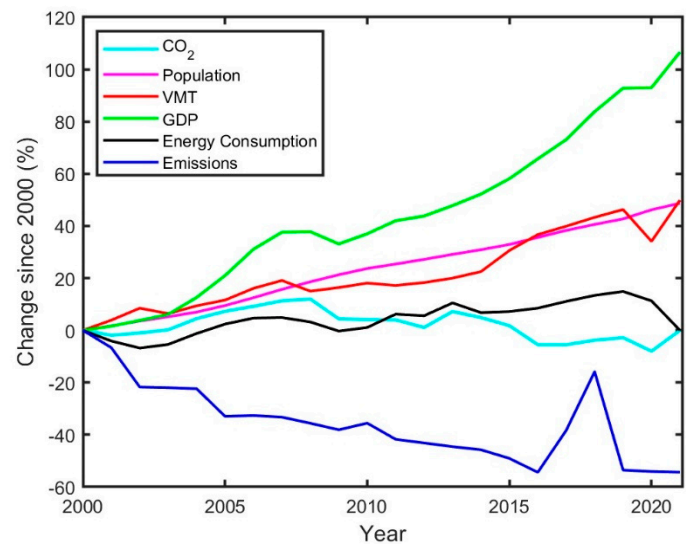

**Figure 1.** Comparison of growth areas and emissions for Utah from 2000–2022. A comparison of GDP [7], VMT [8], population [9], energy consumption [10], $CO_2$ emissions [11], and aggregate emissions [12] normalized to the year 2000.

The first Utah state air quality regulation was passed in 1891, 64 years before the first Federal Act was enacted [13]. This municipal ordinance, enacted in Salt Lake City (UT), mandated the installation of emission control devices called "smoke consumers" and was designed to reduce emissions from coal used in furnaces and boilers. The first federal air act, the Air Pollution Control Act, was declared in 1955 [13]. The first Federal law was the Air Pollution Control Act which was prompted by high-mortality events caused by industrial pollution seen in places such as Meuse Valley, Belgium (1930), Donora, Pennsylvania (1948), and the London Fog event (1952). In 1970, the Environmental Protection Agency (EPA) was established to create a national standard for air quality and to monitor air quality on a national scale. There are two types of National Ambient Air Quality Standards (NAAQS) defined in the Clean Air Act (1970): primary standards, which include health protection, and secondary standards, which include public welfare protection and prevention of decreased visibility. Table 1 shows the years, and regions of Utah, that have been classified as nonattainment areas, regarding meeting the NAAQS standards since 1992.

While a narrative review has been written about air quality in Utah, there is not yet, to our knowledge, a systematic review summarizing the sources and identity of emissions across the state of Utah [13]. The state of Utah, USA, has captured the interest of many researchers not only because of the need for air quality research in the region but also because it possesses unique geographical attributes that can exacerbate air quality issues. The propensity for the state of Utah to experience elevated concentrations of particulate matter and ozone can in part be attributed to its unique geography that features dry, mountainous topography. Valleys in Utah create ideal environments for extended cold-pool events. In this review, we divide the state of Utah into six regions: Utah Valley, Summit County, Southern Utah (The entire southern half of the state, south of Utah Valley (county)), Cache Valley, Uinta Basin, and Salt Lake Valley. This is shown in Figure 2. These regions have been chosen due to their topographical individualities and for easy separation of research area. Admittedly, the reviewed regions do not cover the entire state; however, the reviewed regions cover all the areas in which peer-reviewed air quality research has been conducted. We do acknowledge that other entities have published air quality reports for the State of Utah, however, this review will cover only peer-reviewed literature. A summary of each paper published on outdoor air quality is provided in chronological order for each of the first four regions which have a small number of publications. The last two regions, Uinta Basin and Salt Lake County, have been separated by main species studied for easier reading. Concluding each section is an analysis of what research is missing in each of the six regions.

Table 1. Utah history of nonattainment and maintenance status. Revised from EPA published table [14].

| County | NAAQS | Area | Nonattainment Years | Redesignation to Maintenance | Classification | Whole or/Part County | Population (2010) |
|---|---|---|---|---|---|---|---|
| Box Elder | $PM_{2.5}$ | Salt Lake Valley | 2009–present | - | Serious | Part | 49,057 |
| Cache | $PM_{2.5}$ | Cache Valley | 2009–2020 | 18 June 2021 | Moderate | Part | 112,675 |
| Davis | 8-h Ozone | Salt Lake Valley | 2018–present | - | Marginal | Part | 15,338 |
| Davis | $PM_{2.5}$ | Salt Lake Valley | 2009–present | - | Moderate | Whole | 306,479 |
| Duchesne | 8-h Ozone | Uinta Basin | 2018–present | - | Marginal | Part | 15,338 |
| Salt Lake | 8-h Ozone | Salt Lake Valley | 2018–present | - | Moderate | Whole | 1,029,655 |
| Salt Lake | $PM_{10}$ | Salt Lake Valley | 1992–2019 | 27 March 2020 | Moderate | Whole | 1,029,655 |
| Salt Lake | $PM_{2.5}$ | Salt Lake Valley | 2009–present | - | Serious | Whole | 1,029,655 |
| Salt Lake | Sulfur Dioxide | Salt Lake Valley | 1992–present | - |  | Whole | 1,029,655 |
| Tooele | $PM_{2.5}$ | Salt Lake Valley | 2009–present | - | Serious | Part | 54,857 |
| Tooele | 8-h Ozone | Salt Lake Valley | 2018–present | - | Moderate | Part | 54,857 |
| Tooele | Sulfur Dioxide | Salt Lake Valley | 1992–present | - |  | Part | 58,218 |
| Uinta | 8-h Ozone | Uinta Basin | 2018–present | - | Marginal | Part | 31,979 |
| Utah | 8-h Ozone | Utah Valley | 2018–present | - | Marginal | Part | 515,895 |
| Utah | Carbon Monoxide | Utah Valley | 1992–2005 | 3 January 2006 | Moderate > 12.7 ppm | Part | 166,596 |
| Utah | $PM_{10}$ | Utah Valley | 1992–2019 | 27 March 2020 | Moderate | Whole | 516,564 |
| Utah | $PM_{2.5}$ | Utah Valley | 2009–present | - | Serious | Part | 517,564 |
| Weber | 8-h Ozone | Salt Lake Valley | 2018–present | - | Moderate | Part | 224,583 |
| Weber | $PM_{10}$ | Salt Lake Valley | 1995–2019 | 27 March 2020 | Moderate | Part | 82,825 |
| Weber | $PM_{2.5}$ | Salt Lake Valley | 2009–present | - | Serious | Part | 224,988 |

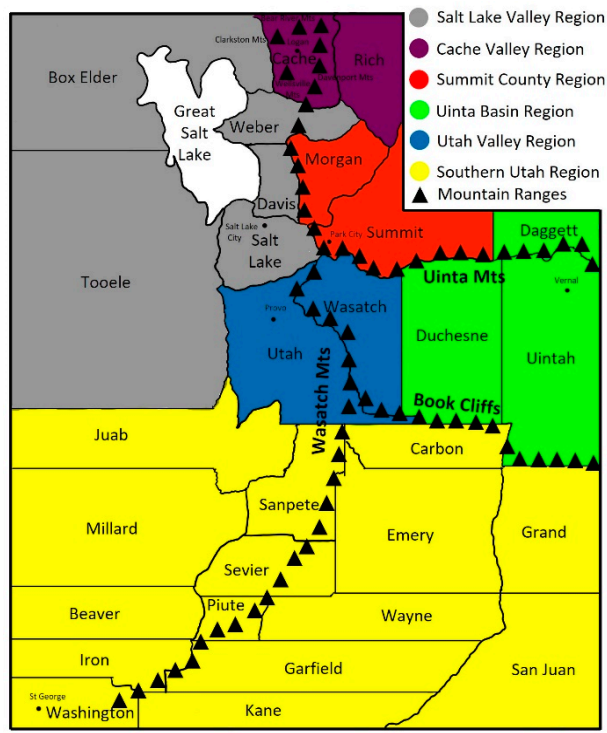

**Figure 2.** Map of Utah, USA with the six regions reviewed in this manuscript. Key cities indicated by black dots.

## 2. Utah Valley

The air quality in this region has been described in three publications between 2002 and 2022. The Utah Division of Air Quality (DEQ) has two active air monitoring sites in the Utah Valley Region (Lindon—490494001, Spanish Fork—490495010) [15]. Utah Valley is situated between two large mountain ranges. As such, it is prone to persistent cold-air pool events which lead to elevated levels of $PM_{2.5}$. This valley also suffers from summertime ozone spikes.

Grover et al. conducted a source apportionment study in Lindon, Utah in August 2002 to determine the composition of $PM_{2.5}$ [16]. Monitoring instruments include a Tapered Element Oscillating Microbalance (TEOM), a Real-time Total Ambient Mass Sampler (RAMS), an ion-chromatographic instrument, an Anderson Aethalometer, and a Particle Concentrator-Brigham Young University Organic Sampling System (PC-BOSS) sampler. Gas phase components were also monitored. The data showed that the average fine particulate composition in the region was composed of 1.3% semi-volatile nitrate, 3.7% non-volatile ammonium nitrate, 5.4% elemental carbon, 9.3% semi-volatile organic carbon, 22.3% ammonium sulfate, and 58.1% non-volatile organic material.

Hansen et al. conducted an air sampling campaign in Lindon, UT, from 19 January to 16 February 2007, designed to measure the mass and composition of urban aerosol on a semi-continuous basis [6]. Their measurements were made at a sampling site located adjacent to the State of Utah Air Quality Monitoring site located at the Lindon Elementary School (EPA AIRS Code 490494001). The following instruments collected $PM_{2.5}$ and $PM_{10}$ mass data: R&P TEOM (nonvolatile PM), R&P filter dynamics measurement system tapered element oscillating microbalance (FDMS TEOM) 8500 (semi-volatile and nonvolatile PM), and Grimm 180 optical particle monitor with a Nafion dryer (semi-volatile and nonvolatile PM). The inorganic composition (sulfate, nitrate, chloride) of $PM_{2.5}$ was determined by the use of an Ambient Ion Monitor (URG 9000 Series). An Andersen Aethalometer was used to measure the contributions of black carbon, diesel, and wood smoke to the $PM_{2.5}$ composition. For the first two weeks of their study, there was a high-pressure system over Utah Valley. The high-pressure system was coupled with snow cover on the ground, high relative humidity, and freezing temperatures. After the first two weeks of monitoring, a frontal passage cleared the inversion and reduced the concentration of all measured species. While other minor inversions did settle into the valley, the concentrations of the species continued to be lower than during the extended cold pool event. It was also noted that a nearby steel mill was not operational during this study and permanently shut down in 2002.

Their measurements showed that most aerosol particulates were comprised of ammonium sulfate and ammonium nitrate salts, the dominant particles being ammonium nitrate with the smaller component being made up of organic material. The peaks in ammonium nitrate containing $PM_{2.5}$ occurred during the afternoons. Hansen et al. suggested that during an inversion, diurnal photochemistry occurred as $NO_x$ precursors were converted to nitrate as a secondary pollutant; mobile sources and wood smoke sources dominated in this region. They further suggested that the evening formation of nitrates could result from the heterogeneous formation of nitrate with increased evening humidity.

The concentration of ammonium sulfate was approximately 1 µg/m$^3$ during the inversion and decreased to 0.5 µg/m$^3$ after the inversion was disrupted. Hansen et al. reported that there were no industrial sources of coal combustion in the valley, a prominent source of ammonium sulfate, and, therefore, determined that the sources of ammonium sulfate must be from mobile sources and wood smoke. More prominent peaks of ammonium sulfate, 2–5 µg/m$^3$, were observed during the inversion event in the late morning and early afternoon, but the source of those spikes remained unknown.

Chloride data showed the most prominent peaks on weekdays during the early morning and late afternoon. As chloride does not have significant precursors to gas emission sources such as nitrate and sulfate, Hansen et al. concluded that the source of chloride must be the sodium chloride used on the roads to prevent icy road conditions. These spikes were consistent with the times of rush hour traffic when cars would be lofting salt from the roads.

Organic carbon measurements were not taken, but total organic material was estimated by subtracting the black carbon, ammonium nitrate, ammonium sulfate, and sodium chloride from the total $PM_{2.5}$ concentrations. This estimate was justified by observing a layer of snow on the ground during the study, ensuring that crustal material was a minor component in the mass of measured $PM_{2.5}$.

The average composition of $PM_{2.5}$ in this study, found in Table 2, was: black carbon 3.9%, ammonium sulfate 1.7%, ammonium nitrate 29%, sodium chloride 1.3%, nonvolatile organic matter 42.5%, and semi-volatile organic matter 21.7%. $PM_{2.5}$ concentrations showed that two-thirds of the $PM_{2.5}$ was carbonaceous, so an apportionment analysis using positive matrix factorization (PMF) was conducted. Hansen et al. identified four primary sources (mobile diesel, mobile gasoline, wood smoke, road dust) and three secondary sources (nitrate, organic matter dominated and organic matter) that contribute to $PM_{2.5}$ concentrations. Sulfate was associated with both groups in the analysis. The average total $PM_{2.5}$ concentration for this period (65.5 µg/m$^3$) had the following composition: gasoline 2.1 µg/m$^3$, diesel 3.4 µg/m$^3$, road dust 4.0 µg/m$^3$, wood smoke 5.3 µg/m$^3$, sulfate 3.5 µg/m$^3$, secondary nitrate 14.7 µg/m$^3$, secondary organic dominated 18.3 µg/m$^3$, secondary organic matter, 16.1 µg/m$^3$. The apportionment analysis also showed that two-thirds of the $PM_{2.5}$ was due to the secondary formation of ammonium nitrate and organic material, the precursors being identified as $NO_x$ and volatile organic compounds emitted from mobile emissions and wood smoke. Hansen et al. concluded that diurnal homogenous photochemical and heterogeneous nighttime processes were involved in making a large portion of the $PM_{2.5}$ in Utah Valley. Hansen et al. also concluded that a large amount of the $PM_{10}$ was from vehicles liberating particles from the road as $PM_{10}$ followed a similar trend to chloride concentrations and had a large chloride contribution.

Cropper et al. improved on the study by Hansen et al. by using a gas chromatography-mass spectrometry organic aerosol monitor (GC/MS OAM) deployed on the Brigham Young University (BYU) campus from January through February 2015 [17]. Cropper et al. used a GC/MS OAM to identify organic marker compounds for a PMF analysis, including lev-oglucosan, dehydroabietic acid, stearic acid, pyrene, and anthracene. An FDMS TEOM monitored $PM_{2.5}$, and an ambient ion monitor (AIM) was used to measure particulate-phase anions and cations. Cropper et al. recorded that the only ions with significant concentrations were nitrate, sulfate, and ammonium. An aethalometer monitored black carbon in $PM_{2.5}$. $NO_x$, ozone, and CO were also monitored. The PMF analysis that included fifteen species yielded nine significant factors which are also shown in Table 3: nitrate 6.81 µg/m$^3$, diesel 0.405 µg/m$^3$, anthracene 0.16 µg/m$^3$, wood smoke 5.05 µg/m$^3$, auto 0.42 µg/m$^3$, organic material 3.51 µg/m$^3$, ozone 0.52 µg/m$^3$. Cropper et al. concluded that 34% of the $PM_{2.5}$ was from primary sources including wood smoke, diesel, anthracene, and auto factors, with wood smoke being the dominant factor. Sixty-six percent of $PM_{2.5}$ came from secondary formation and was dominated by secondary nitrates. Cropper et al. also identified the composition of wood smoke as 0.3% ammonium sulfate, 5.8% ammonium nitrate, 2.5% dehydroascorbic acid (DHAA), 7.4% levoglucosan, 4% black carbon, and 80% organic matter.

**Table 2.** Summary of Hansen et al. source apportionment results.

| Factor | Makeup of $PM_{2.5}$ (%) |
| --- | --- |
| Black carbon | 3.9 ± 9 |
| Ammonium sulfate | 1.7 ± 9 |
| Ammonium nitrate | 29.0 ± 9 |
| Sodium chloride | 1.3 ± 9 |
| Semi-volatile organic material | 21.7 ± 20 |
| Nonvolatile organic material | 42.5 ± 20 |

**Table 3.** Summary of Cropper et al. source apportionment results.

| Factor | Makeup of $PM_{2.5}$ (%) | Makeup of Wood Smoke (%) | |
| --- | --- | --- | --- |
| Nitrate | 40.4 | Levoglucosan | 7.4 |
| Diesel | 2.4 | dehydroascorbic acid | 2.5 |
| Anthracene | 0.9 | Ammonium sulfate | 0.3 |
| Wood smoke | 29.9 | Ammonium nitrate | 5.8 |
| Auto | 2.5 | Black carbon | 4.0 |
| Organic Material | 20.8 | Organic material | 80 |
| Ozone | 3.1 | | |

*Summary of Utah Valley Region*

In Utah Valley, ammonium nitrate is the dominant component of $PM_{2.5}$, with ammonium sulfate contributing between 2–5 µg/m$^3$ when $PM_{2.5}$ spikes to concentrations >65.5 µg/m$^3$. These spikes exceed the EPA standard for $PM_{2.5}$ which skews the 24-h average used by the EPA to define nonattainment. Nitrates and organic material are the leading anthropogenic emissions in the valley, contributing 91.1% of the mass of $PM_{2.5}$. While source apportionment studies have made significant progress and provided an understanding of the state of air quality in Utah Valley, a more complete survey of pollutants in Utah Valley has not yet been completed. The pollutants mentioned in the NAAQS standards, such as Pb and $SO_2$, have not yet been studied in Utah Valley but more importantly speciated volatile organic compounds (VOCs) have not been measured. Measurement of speciated VOCs would aid in understanding the principal drivers of ozone formation and contribute to a better understanding of secondary organic aerosol (SOA) production in this region. Long-term monitoring studies of ozone and dust sources have not yet been published.

## 3. Summit County

The air quality in this region has been described in one publication between 2002 and 2022. The Utah DEQ has no active air monitoring sites in the Summit County Region [15].

Mendoza et al. conducted a study on fine particulate matter in Park City from 3 February 2020, through 23 July 2020 [18]. This article studies the trends in $PM_{2.5}$ during the corona virus pandemic of 2020 when much of the world, including Summit County, went through a period of mandatory quarantine. During the quarantine, or "Lockdown", people were encouraged to remain at home where possible. It should be noted that the "Lockdown" period in Utah began on 16 March and ended on 26 April 2020. Mendoza et al. concluded that at the commercial site, during the lockdown and easing periods, $PM_{2.5}$ decreased. The Reopening period, 12 June–23 July 2020, marked a rise in $PM_{2.5}$ relative to the lockdown period, 16 March–26 April 2020, and Easing periods, 1 May–11 June 2020, but it was still lower than the pre-lockdown period, 3 February–15 March 2020, levels. These data can be found in Table 4. The same decrease was seen during the lockdown and easing periods at the residential site. During this time, evening peaks started at 6 pm and persisted until midnight which, Mendoza et al. contributed to wood-burning fires that could be detected from the sampling site.

Table 4. Summary of Mendoza et al. average $PM_{2.5}$ findings.

| | Commercial (µg/m$^3$) | | Residential (µg/m$^3$) | |
| --- | --- | --- | --- | --- |
| **Time Period** | **Weekdays** | **Weekends** | **Weekdays** | **Weekends** |
| Pre-lockdown | 2.40 | 1.95 | 2.35 | 1.40 |
| Lockdown | 1.30 | 1.30 | 1.23 | 1.33 |
| Easing | 1.38 | 1.13 | 1.40 | 1.58 |
| Reopening | 2.20 | 2.13 | 2.30 | 2.20 |

Mendoza et al. concluded that the diurnal atmospheric stagnation they observed resulted in the secondary formation of $PM_{2.5}$ during this study. However, Mendoza et al. also concluded that the cold air pools observed in Park City led to less secondary $PM_{2.5}$ formation than an equivalent cold air pool formed at lower altitudes. While meteorological conditions may have affected $PM_{2.5}$ concentrations, a rise in concentrations back to pre-lockdown concentrations during the reopening period suggests that meteorology did not significantly affect the measured concentrations.

*Summit County Summary*

Summit County has been neglected in air quality studies. The study by Mendoza et al. is the only published air quality paper in the database. This study covered a period of measuring $PM_{2.5}$, but more extensive studies of $PM_{2.5}$, including source apportionment, should be conducted under normal conditions as these were executed mostly under pandemic conditions. It should be noted that the changes in $PM_{2.5}$ concentrations are too low to extract more comprehensive conclusions from the data. The reported $PM_{2.5}$ levels do, however, fall within the EPA guidelines. Studies of dust sources, CO, Pb, $NO_x$, ozone, $PM_{10}$, and $SO_2$, are also lacking for this region. Organic compounds, such as VOCs, and black and elemental carbon sources have not been measured. No PMF analysis or source apportionment studies have been conducted on the Summit County side of the Wasatch mountains. In short, little is known about air quality in Summit County and the east side of the Wasatch Mountains.

## 4. Southern Utah

The air quality in this region has been described in three publications between 2002 and 2022. The Utah DEQ has four active air monitoring sites in the Southern Utah Region (Carbon—490071003, Garfield—490170006, Iron—490210005, Washington—490530007) [15]. The Southern Utah region is largely desert-like terrain.

From March 2011 to March 2012, Hahnenberger et al. conducted a dust composition study on the playa of the Sevier Dry Lake using BSNE dust monitors [19]. Dust and soil samples were collected, resuspended, and resolved to measure the elemental composition of dust particles. Most elements were found to have a significant enrichment factor in the dust relative to the soil, including Na, P, S, Cl, Ti, V, Cr, Ni, Cu, Zn, Ga, Se, Br, Rb, Sr,

Y, and Zr. However, the soil was more enriched than the dust in Si, K, Ca, Mn, and Fe, which are also the major soil compositional elements. The most significant enrichment factors in dust were found for Zn, Se, Br, and Y, all minor soil compositional elements. The reported enrichment factors compared the elemental concentrations found in dust particles to the average values for each element calculated over the entire crustal mantle of the earth, not just for average crustal values found in the Utah crust. For all the elements that had significant enrichment factors, it was found that the fine particles had more significant mean enrichment factors than the coarse particles. However, it was noted that most of the major elements did not have significant differences in concentration between the fine and coarse particles. The exceptions were Ca, Mg, and Na, in the course particles and in fine particulates Si and Ti. All these elements are significant in soil composition except for Na; however, Na is present in the lake due to its association with sodium salts and minerals. High levels of Cr, As, Pb, and Ni are known to have detrimental health effects; however, these were not found to be a concern in the dust from this playa.

Reynolds et al. conducted a mineral aerosol analysis that stretched across the central Rocky Mountain region, part of which extends into southern Utah [20]. A portion of the study took place in the Canyonlands National Park (CNP), UT. Other locations investigated took place in surrounding states to Utah and, therefore, are not discussed in this review. A total suspended particulate (TSP) study was executed to track the movement of particles in this region. It was found that fine particles travel in this region and their movement is affected by wind, fires, dust storms, snowpack, monsoons, and other meteorological events. This study took place over two years in CNP, and the reported mean results included TSP (Thermo Finnigan GS2310 High Volume Flow Controller, Thermo G313 Mass Flow Controller), $PM_{2.5}$ (Malvern Mastersizer), $PM_{10}$ (Malvern Mastersizer), organic material (OM) (loss-on-ignition analysis on filters), and organic material percentage in the daily TSP readings. These results are reported in Table 5.

**Table 5.** Mean concentrations for all recorded data. All values are reported in $\mu g/m^3/day$, sd is the standard deviation of the data.

| Year | TSP (sd) | $PM_{2.5-10}$ | $PM_{2.5}$ | Organic Material | Organic Material (%) |
|---|---|---|---|---|---|
| 2011 | 134 (75) | 3.8 (2.3) | 2.5 (0.8) | 3.1 (0.7) | 3.7 |
| 2012 | 171 (111) | 4.6 (3.0) | 3.2 (0.8) | 3.4 (0.8) | 2.8 |

Hall et al. measured the nitrogen-containing ion species in $PM_{2.5}$ in Southern Utah. The results are found in Table 6.

**Table 6.** Annual average concentrations of nitrogen species measured by Hall et al. [21].

| Species | Green River (mgNL$^{-1}$) | Great Basin (mgNL$^{-1}$) | Pinedale (mgNL$^{-1}$) | Craters of the Moon (mgNL$^{-1}$) |
|---|---|---|---|---|
| $NO_3^-$ | 0.29 | 0.16 | 0.10 | 0.08 |
| $NH_4^+$ | 0.38 | 0.26 | 0.12 | 0.20 |

*Southern Utah Summary*

Southern Utah encompasses 66% of the total area of the state of Utah, much of which is rural (small cities and farmlands), United States Forest Service, Bureau of Land Management, state and national parkland, or tribal lands. As such, a minimal number of air quality studies have been executed in southern Utah. The State of Utah does have four DEQ monitoring stations located in Southern, UT, (County with EPA AIRS Code: Carbon—490071003, Garfield—490170006, Iron—490210005, Washington—490530007) and the three Interagency Monitoring of Protected Visual Environments (IMPROVE) sites that exist in the National Parks in Southern Utah [22]. While dust studies have been conducted on the Sevier Dry Lake, no studies have been conducted to see how far this dust travels or where it deposits. PM has been reported as a function of each monitoring station; however, no general analysis or source apportionment study has been executed. Studies of CO, Pb, $NO_x$, TSP, ozone, black and elemental carbon, and $SO_2$ are lacking in this vast region leading to the conclusion that additional information about the sources and sinks of PM can be learned about the air quality of Southern Utah.

## 5. Cache Valley

The air quality in this region has been described in six publications between 2002 and 2022. The Utah DEQ has one active air monitoring site in the Cache Valley Region (Smithfield—490050007) [15]. Cache Valley (1430 m above sea level) is in a geographical bowl surrounded by the Clarkston (2513 m), Wellsville (2857 m), Davenport (2601 m), and Bear River Mountains (3042 m). This bowl makes Cache Valley an ideal environment for extended cold-pool events which leads to elevated levels of $PM_{2.5}$.

In 2004, Malek et al. investigated the claims that Logan, UT, had the worst air quality in the nation [23]. Using air quality data measured at the monitoring station maintained by Utah DAQ in downtown Logan (EPA AIRS Code 490050007), coupled with meteorological readings measured at the Logan, UT airport, Malek et al. categorized $PM_{2.5}$ formation into meteorological factors, air pollutant sources, and formation factors.

Malek et al. cited the US Air Quality Gradebook for emissions data in Cache Valley and concluded that during the winter season, the top pollutants are motor vehicles, wood stove combustion, and agricultural ammonia, which drives both primary and secondary air pollutants. PM builds up in Cache Valley without major precipitation, winds, or in the presence of high-pressure systems and inversions. Malek et al. theorized that the large numbers of cattle, hogs, and sheep located in the region are the primary sources of ammonia. Malek et al. also theorized that vehicle emissions are higher than usual due to cars idling before being used in the winter. Photochemistry occurs in the fog and drives the secondary formation of $PM_{2.5}$. This is especially troublesome in Cache Valley because of the shape of the valley. Cache Valley (1430 m above sea level) is in a geographical bowl surrounded by the Clarkston (2513 m), Wellsville (2857 m), Davenport (2601 m), and Bear River Mountains (3042 m).

In January 2004, Cache Valley had 392 mm of snow coverage. The valley fluctuated between low and high pressures, often separated by winds and precipitation. During the high-pressure periods, $PM_{2.5}$ reached 132.5 µg/m$^3$ for a 24-h average. Malek et al. state that the snow covering the valley floor was responsible for reflecting up to 80% of incoming solar radiation resulting in more photochemical reactions linked to the secondary formation of $PM_{2.5}$. Malek et al. also noted that high ground-level $PM_{2.5}$ levels indicate inversion layers. Strong inversions were evidenced by high pressures, low surface temperatures, and strong negative net radiation (reflection of radiation). Inversion heights during January 2004 were measured to be up to 350 m. The atmosphere above the inversion was classified as Pasquill–Gifford Stability Class G, meaning that the measured ambient lapse rate was less than +4 °C/100 m and that vertical mixing was negligible under these conditions. Thus, Malek et al. concluded that not only are the air pollutant sources causing high levels of $PM_{2.5}$ but topographical conditions existing in Cache Valley contribute to $PM_{2.5}$ formation.

Silva et al. studied the same January 2004 inversion in Cache Valley [4]. They used a quadrupole aerosol mass spectrometer (Q-AMS) to determine the chemical composition of $PM_{2.5}$. During this period, it was found that the varying percentage of total mass concentration of $PM_{2.5}$ was as follows: nitrate 40–53%, sulfate 4–7%, ammonium 10–18%, organic carbon 11–22%, elemental carbon 1–3%. If the inversion remained stagnant for several days, spikes of ammonium nitrate could rise to 80–85% of the total mass concentration of $PM_{2.5}$. The partitioning constant, Kp, at that time, was 0.001 ppb$^2$ due to the low surface temperature. Gas-phase ammonium mixing ratios were measured to be 14 ppb in downtown Logan and up to 370 ppb in rural Cache Valley for an hourly average, both of which are above the partitioning constant, meaning that $PM_{2.5}$ was being made in the area as opposed to being transported to the region. It was calculated that ammonium nitrate formation was limited by nitric acid during the winter. This was confirmed by ammonia gas and particle concentrations being 2–3 times greater than the total nitrate concentrations in the valley. It was unknown whether nitric acid was formed from $NO_x$ or VOC sources. The theories of the sources of pollutants were the same as Malek et al.

Wang et al. quantified the frequency and intensity of inversions in Cache Valley for 57 years and used these data to develop a relationship between the occurrence of inversions, mean surface air temperature and the likely concentration of $PM_{2.5}$ reached during extended cold pool events [5]. This was then used to approximate $PM_{2.5}$ in earlier periods based on the relationship between inversion intensity, duration, and air temperature results. They found seven categories for the likelihood of an inversion based on the $\Delta T$ of the day (Tmax − Tmin in °C). The higher the $\Delta T$, the higher the category, and the higher the chance of an inversion. A regression model of log($PM_{2.5}$) was explored with five parameters ($\Delta T$, mean air temperature, inversion duration, and snow cover at two sites) to produce an equation used to model the data prior to $PM_{2.5}$ being directly measured in this area. The results from this model suggest an increase in $PM_{2.5}$ in Cache Valley in the 1980s that declined after 1990.

From December 2013 to February 2015, Hall et al. collected precipitation samples to measure nitrogen deposition concentrations [21]. These samples were taken within 36 h from the end of the precipitation event to minimize bacterial processing of the nitrogen. The samples were analyzed for $NO_3^-$ and $NH_4^+$, total dissolved nitrogen, and nitrogen isotopes. Dissolved organic nitrogen was calculated from the total dissolved, inorganic nitrogen values and nitrogen isotope ratios were also calculated for $PM_{2.5}$. It was found that the total bulk N deposition and annual concentrations, found in Table 7, of $NH_4^+$ and $NO_3^-$ did not significantly differ between the urban (Salt Lake) and agricultural (Cache) valleys in this study. It was found that $NO_3^-$ did differ seasonally, specifically at the height of summer in June and July, between the valleys: $0.65 \pm 0.10$ mgNL$^{-1}$ in Salt Lake Valley and $0.42 \pm 0.08$ mgNL$^{-1}$ in Cache Valley. Some patterns and differences were found in $NH_3$ and $NH_4^+$ that suggest local agricultural impacts on emissions are only seen on finer spatial scales rather than across the entire valley. This was shown by the differences between rural and urban testing sites in Cache Valley. Patterns were also affected by plants' removal of atmospheric N, which mainly affects bulk deposition concentrations. This was evidenced by the summer canopy observed in Cache Valley, which is not prevalent in Salt Lake Valley, which showed a greater net plant uptake of $NH_3$ and $NO_2$. Hall et al. concluded that differences in the mean speciation, isotope composition, or bulk concentrations were not observed. The only differences between the valleys were found in the summer months and were seen in differences in $NO_3^-$ and $^{15}N$ concentrations, which were higher in the Salt Lake Valley than in Cache Valley.

Table 7. Annual concentrations of nitrogen species measured by Hall et al. [21].

| Species | Cache | Salt Lake |
| --- | --- | --- |
| Ammonium | $0.51 \pm 0.04$ mgNL$^{-1}$ | $0.45 \pm 0.02$ mgNL$^{-1}$ |
| Nitrate | $0.39 \pm 0.04$ mgNL$^{-1}$ | $0.45 \pm 0.03$ mgNL$^{-1}$ |
| $NO_2$ | 11 ppb | 14 ppb |
| Inorganic N | $0.14 \pm 0.05$ mgNL$^{-1}$ | $0.14 \pm 0.04$ mgNL$^{-1}$ |
| Cumulative Bulk N Deposition | 3.5–5.1 kg N ha$^{-1}$ yr$^{-1}$ | 3.5–5.1 kg N ha$^{-1}$ yr$^{-1}$ |

Franchin et al. conducted airborne and ground-based observations of ammonium nitrate aerosols from 16 January 2017 to 12 February 2017 [24]. Airborne and ground observations were made in the Cache, Salt Lake, and Utah Valleys. All three valleys showed similar airborne trends and mass concentrations that increased with time. During these studies, 70% of the measured $PM_{10}$ was $PM_{2.5}$, and three pollution episodes were identified. During the first episode, Cache Valley had the highest $PM_{2.5}$ concentrations (airborne 70 µg/m$^3$ and ground 90 µg/m$^3$), and Utah Valley had the lowest (airborne 10 µg/m$^3$ and ground 25 µg/m$^3$). In the second period, data were collected at ground sites in Cache Valley (100 μg/m$^3$), Utah Valley (70 μg/m$^3$) and Salt Lake Valley (60 μg/m$^3$). Franchin et al. defined ≤2 μg/m$^3$ as clean air and ≥17.5 μg/m$^3$ as polluted conditions. During polluted conditions, ammonium nitrate was the dominant species being 74 ± 5% (mean and standard deviation), while organic species (18 ± 3%), sulfate (6 ± 3%), and chlorides (2 ± 2%) constituted smaller fractions of the total mass concentrations. The clean regions between pollution periods saw that the mass of aerosols was dominated by an organic fraction (50 ± 13%), followed by ammonium nitrate (31 ± 9%), ammonium sulfate (13 ± 7%), and chloride (6 ± 6%). Pollution episode two saw an average ground increase of 11 μg/m$^3$ day$^{-1}$. The 11 μg/m$^3$day$^{-1}$ value can be deconvolved into 8.2 ± 0.4 μg/m$^3$day$^{-1}$ from ammonium nitrate and 2.8 ± 0.1 μg/m$^3$day$^{-1}$ from other species. This change was uniform across all three valleys by airborne measurements; however, ground measurements were only confirmed in Cache Valley. A large plume of US Magnesium drifted across the area and resulted in an increase in the organic matter, chloride and a slight rise in ammonium found in the PM$_{2.5}$. It was also found that 20% of the total organic material in the sampled n aerosols during pollution periods was levoglucosan which is linked to wood smoke. Secondary aerosols and other organic aerosol components were not traced. Average vertical profiles showed that 75% of the aerosol mass is ammonium nitrate at 300 m above ground level (a.g.l) in Cache Valley, 500 m a.g.l in Salt Lake Valley and 300 m a.g.l in Utah Valley. Above those cut-offs, there is variability in aerosol composition. Nitric acid was the limiting reagent in particulate formation in all three valleys. A simulation, using the ISORRIPIA model, was conducted to see what a total decrease of 50% in ammonium would do to the rate of aerosol production. Input parameters for the model were obtained using the Twin Otter aircraft and a high-resolution time-of-flight iodide adduct ionization mass spectrometer (HRToF-CIMS). The model predicted a 44% decrease in PM$_{2.5}$ Salt Lake Valley, and a 37% decrease in Utah Valley (37%) but no decrease was predicted for Cache Valley. When the same simulation was run with a 50% decrease in nitrate, there was a 44% decrease in PM$_{2.5}$ across all three valleys.

Moreover, from January to February 2017, Mukerjee et al. from the EPA made measurements of PM, non-VOC gaseous compounds, VOCs, and meteorological conditions [25]. Their measured 12-h averaged concentrations are found in Table 8. The 1-h averages were also given in their paper. The Wilcoxon rank-sum test was used and no statistically significant differences in PM or VOCs at nighttime compared to daytime were observed; however, NO$_2$ was higher at night. These findings, coupled with the calculation of Spearman correlation coefficients, showed that this was expected based on the lack of photochemistry during nighttime hours. The Spearman correlation coefficients were positive when correlating nitrogenous pollutants with PM, and ozone was negative when correlated with all other pollutants. PM and VOCs were positive when correlated with other PM or VOC species, and VOCs were strongly correlated with PM and most gaseous pollutants. NO$_x$ was seen to increase in the mornings and decrease in the afternoons, followed by a spike in ozone in the late mornings. This is consistent with known photochemical processes. Hour-of-the-day average PM concentrations were found to be in the range of 22–28 μg/m$^3$. The Kruskal–Wallis test found that most pollutants were lower in concentration when the wind blew from a northern direction. The one exception is ozone which rose in concentration when the wind blew from a westerly direction. It was also observed that PM$_{2.5}$ and ozone concentrations were lower when winds blew from a southeastern direction.

*Cache Valley Summary*

More research has been executed in Cache Valley than in the other regions discussed so far. Much has been undertaken to analyze air quality during inversions in Cache Valley, but more could be accomplished to monitor normal baseline conditions throughout the year. Inversions are particularly strong in Cache Valley due to the topography of the mountains in this region. Cache Valley has mountains on all four sides which create a bowl-like valley making an ideal environment for extended cold-pool events. No source apportionment studies have been executed of the PM$_{2.5}$ measured in the region. The composition of PM$_{2.5}$ has, however, been determined. NO$_2$ studies have been conducted in the area but should be expanded. While it is important to study PM$_{2.5}$ in this region as this is what the region struggles most with, it would be prudent to expand studies to other species in the NAAQS standards, including O$_3$, CO, Pb, SO$_2$, and PM$_{10}$. A design and rationale have been compiled to more comprehensively study Cache Valley air quality, as well as Utah and Salt Lake Valleys, which displays the need for further study and outlines possible procedures that could be used to conduct these studies [26].

**Table 8.** 12-h averages of pollutants and meteorological variables 16 January–16 February 2017. NO$_x$ was calculated as the sum of NO$_2$ and NO, and NO$_z$ as the difference of NO$_y$–NO$_x$.

| Pollutant/Meteorological | Median | Mean | Minimum | Maximum |
|---|---|---|---|---|
| NO$_2$ (ppb) | 14.2 | 16.6 | 2.7 | 40.8 |
| NO (ppb) | 2.9 | 7.4 | 0.0 | 36.4 |
| NO$_x$ (ppb) | 20.0 | 24.1 | 2.9 | 63.1 |
| NO$_y$ (ppb) | 24.8 | 29.4 | 1.7 | 83.0 |
| NO$_z$ (ppb) | 3.7 | 5.7 | 0.1 | 21.3 |
| O$_3$ (ppb) | 22.7 | 21.3 | 0.8 | 41.2 |
| PM$_{10}$ (µg/m$^3$) | 24.7 | 34.6 | 2.2 | 119.4 |
| PM$_{2.5}$ (µg/m$^3$) | 16.3 | 24.0 | 1.2 | 88.3 |
| PM$_{10-2.5}$ (µg/m$^3$) | 7.6 | 10.6 | 1.0 | 31.1 |
| Temperature (°C) | −2.2 | −2.9 | −15.3 | 7.4 |
| Humidity (%) | 79.3 | 77.6 | 52.2 | 90.4 |
| Pressure (mbar) | 860.1 | 860.1 | 837.7 | 880.0 |
| Wind speed (m/sec) | 1.1 | 1.3 | 0.5 | 4.2 |
| Total precipitation (mm) | 0 | 0.6 | 0 | 14.8 |
| Mixing height (m) | 236.7 | 541.3 | 160.5 | 1693.8 |
| Benzene (ppbC) | 1.64 | 1.91 | 0.63 | 4.04 |
| Toluene (ppbC) | 4.46 | 5.75 | 0.77 | 15.11 |
| Ethylbenzene (ppbC) | 0.71 | 0.87 | 0.17 | 2.20 |
| m,p-xylene (ppbC) | 2.54 | 3.03 | 0.61 | 7.54 |
| o-xylene (ppbC) | 1.06 | 1.27 | 0.21 | 3.91 |
| Methyl chloride (ppbC) | 1.00 | 1.06 | 0.74 | 2.57 |
| Toluene/Benzene (ppbC) | 2.67 | 2.96 | 0.41 | 10.42 |
| m,p-xylene/ethylbenzene (ppbC) | 3.61 | 3.46 | 0.80 | 6.24 |
| Sum of xylenes/benzenes (ppbC) | 2.21 | 2.44 | 1.20 | 3.50 |

## 6. Uinta Basin

The air quality in this region has been described in twenty-one publications between 2002 and 2022. The Utah DEQ has two active air monitoring sites in the Uinta Basin Region (Vernal—490471004, Roosevelt—490130002) [15]. An interesting anomaly regarding the Uinta Basin is that the Uinta Basin is a region that is prone to intense thermal inversions. This is abnormal as most intense ozone events occur in summer when maximum solar radiation occurs and in large cities that have elevated concentrations of ozone and ozone precursors from anthropogenic emissions [27]. The Uinta Basin is known to have persistent snow cover and is a region where fossil fuel extraction is prominent. Ozone formation in the Uinta Basin correlates strongly with thermal inversions and snow cover [27]. These inversions allow for the accumulation of ozone precursors under a tight boundary layer and increased surface albedo created by the accumulation of snow cover intensifies the amount of solar energy available for ozone production.

*6.1. Ozone Studies*

In 2012, Edwards et al. conducted a six-week sampling campaign in the Uinta Basin [28]. The Uinta Basin is known to have high levels of ozone which have been linked to the oil and gas production in this region. Edwards et al.'s air sampling campaign yielded an observation of a diurnal cycle for ozone. Edwards et al. concluded that the nighttime decrease was due to NO$_x$ reaction pathways and the daytime increase was due to photochemical production. Their simulations observed a decrease in VOCs correlated with an increase in ozone. The simulation also showed that OH formation from O1(D)—from ozone photolysis, reacting with water vapor only accounted for 170 pptv/day$^{-1}$ of OH, accounting for 7.6% of primary radicals. Formaldehyde was responsible for forming 52.3% of radicals, with HONO (25.8%) and ClNO$_2$ (12.8%) following. An additional 1.5% of radicals were counted as OH radicals formed from the reaction of alkenes with ozone. Aromatic VOCs were 30–40 times more efficient in ozone formation than non-aromatic VOCs, and NO$_x$ concentrations were much higher than radicals. This observation concluded that radicals are the limiting factor in O$^3$ formation in these reactions. Edwards et al. concluded that the system was highly VOC sensitive seeing a ratio of 230 ppbv non-methane hydrocarbons to 5.6 ppbv NO$_x$ over six weeks. Manipulation of the NO$_x$ concentrations in the simulation confirmed these hypotheses.

Other models for the Uinta Basin were made by Mansfield et al. in 2013, specifically for the analysis of winter ozone events [27]. The two models, Ouray-5 and Boulder-6, use lapse rate, snow depth, solar angle, temperature, and the number of consecutive days under inversion details, with the addition of wind speed for Boulder-6, to create regression models that estimate the total number of days in a season that the NAAQS standard for ozone will be exceeded. With enough input, the model was found to have an accuracy of ±1 day and standard errors of 10 and 5 ppb of ozone for each sub-model. Historical data were used to test the model. Ozone events were found to be most intense in February and March, and the one-to-two-month lag could be explained by the intensifying of solar radiation and elevation. It was also found that ozone can peak in relatively weak inversions, not only during intense and prolonged inversions.

Modeling the effects of carbonyl photolysis in the Uinta basin was accurate with some amenable conditions in a box model [29]. The model was compared to data from a six-day sampling campaign from 31 January to 5 February 2013. This was a time of high ozone production with daily spikes of up to 95 ppbv and a daily 8-h average of up to 107 ppbv. The model matched the measurements on a 10-min average with a discrepancy of ±4%. Accurate concentrations of oxidized nitrogen species and oxygenated VOCs were calculated using the model, which gives confidence in the model's ability to simulate VOC-NO$_x$ photochemistry. Despite an investigation into OH production, the most dominant radical source was found to come from the photolysis of carbonyl compounds in the Uinta Basin (85%). A description of the model can be found in their paper [29].

Neeman et al. ran simulations for the Unita Basin, with an emphasis on cold-air pools affecting ozone concentrations [30]. The data used were collected from 31 January to 6 February 2013. The Congestion Mitigation and Air Quality Improvement (CMAQ) model used correlated ozone production to snow cover and cloud microphysics. The characteristics of the cold-air pool under the boundary layer (2100 m) were more susceptible to changes from snow cover than to cloud microphysics. When clean air was blown into the valley, pollutants were mixed. Model estimates of ozone were better during the day than at night and in high precursor condition areas. Table 9 shows the results of their simulation:

**Table 9.** CMAQ generated ozone concentrations by FULL and NONE simulations. FULL assumed snow in the basin and NONE assumed no snow below 2000 m in the basin. [30].

|  | FULL | NONE |
| --- | --- | --- |
| Highest mean ozone afternoon (ppb) | 97.2 | 81.2 |
| Highest mean ozone non-afternoon (ppb) | 61.9 | 51.0 |
| Max hourly ozone (ppb) | 134.4 | 118.0 |
| Area of mean afternoon ozone (km$^2$) | 896 | 114 |

Ahmadov et al. also investigated ozone production in the Uinta Basin through the use of modeling. The data used were from the winters of 2012 and 2013 [31]. Simulations were run using bottom-up and top-down emission estimates. It was found that the Weather Research and Forecasting Chemistry (WRF-Chem) model accurately simulated meteorological conditions and the transport of radicals in the basin. It was found that the 3D meteorological-chemistry model framework and top-down emission estimates worked well and could be used in future simulations in the Uinta Basin.

Three field campaigns during the winters of 2012, 2013, and 2014 were conducted to see how the high wintertime ozone levels in the Uinta Basin affect the nitrogen budget in the area [32]. Ozone and reactive nitrogen were found to be 2.5 times higher in 2013 than in 2012 with the averages for 2014 being in the middle. Photochemically active nitrogen sources ($NO_x$) remained consistent throughout the three years. Large concentrations of $HNO_3$ were found which resulted in particle formation.

An instrumented aircraft flew over the Uinta Basin in February 2012 and again between January and February 2013 to measure the composition of air in the region [33]. The altitude of the aircraft was varied throughout the flight in order to obtain vertical profiles of air composition. In 2012 the $O_3$ mole fraction ranged between 30–45 ppb throughout the basin. Variations were only seen at the peaks that surround the basin. $CH_4$ varied across the basin and varied across subsequent flights. Peaks were seen to be around the gas fields and trailed off in the direction of that day's prevailing winds. In 2013 high levels of $O_3$ were observed in the surface area of the basin. A strong temperature inversion was observed up to 1650 m above sea level (a.s.l.) and a cold-pool layer was observed up to 2200 m a.s.l. Emissions from a powerplant in the eastern region of the basin were observed to be emitted at a higher altitude than that of the ozone maximum in the boundary layer. The authors concluded that the sources of $O_3$ enhancement must already exist in the basin. There was a large difference in snow cover in the basin between 2012 and 2013 and the differences in ozone production were attributed to the difference in ground snow cover.

Schnell et al. did further studies on the 2012 to 2013 winter in the Uinta Basin [34]. They quantified the daytime growth rate of ozone (Table 10), which averaged between 5–11 ppbv/h, and the nighttime loss rate (Table 11), which averaged 1–2 ppbv/h, in the region. The nighttime decrease was attributed to $NO_x$ production from vehicle traffic in the region. The cold pools were measured to be below the rim of the basin meaning that the cold pool did not spill over and out of the basin. Production of ozone reached concentrations of 100 ppbv. Wind analysis showed that the air moves towards the slopes of the basin during the day and back down into the bowl at night.

**Table 10.** Average hourly ozone growth rates in the Uinta Basin [34].

| Level (m) | 31 January 2013 | 1 February 2013 | 2 February 2013 | 3 February 2013 | 4 February 2013 | 5 February 2013 | 6 February 2013 |
|---|---|---|---|---|---|---|---|
| | | | Ouray Growth Rates (ppbv/h) | | | | |
| 2 | 6.0 | 8.3 | 4.4 | 6.1 | 7.0 | 9.2 | 11.2 |
| 50 | 4.0 | 7.7 | 4.5 | 5.9 | 6.5 | 9.2 | 13.0 |
| 100 | 3.8 | 7.3 | 4.3 | 4.9 | 6.1 | 8.9 | 13.0 |
| 150 | 3.6 | 7.3 | 5.7 | 1.9 | 4.1 | 8.7 | 12.6 |
| 200 | 2.9 | 2.6 | 5.8 | −0.9 | 3.0 | 6.8 | 11.9 |
| 250 | 1.8 | 5.2 | 4.5 | 0.5 | 3.8 | 2.1 | N/A |
| | | | Fantasy Canyon Growth Rates (ppbv/h) | | | | |
| 2 | 9.0 | 9.0 | 6.9 | 7.4 | 7.6 | 9.3 | 11.6 |
| 50 | 5.6 | 8.2 | 6.6 | 7.2 | 6.8 | 8.7 | 11.6 |
| 100 | 10.3 | 7.1 | 6.5 | 1.7 | 6.2 | 10.7 | 13.2 |
| 150 | 3.3 | 6.2 | 6.8 | 0.8 | 4.0 | 7.6 | 11.3 |
| 200 | 3.8 | 6.9 | 7.7 | 2.4 | 2.2 | 12.0 | 11.1 |
| 250 | 2.0 | 3.7 | 8.8 | 3.1 | 4.1 | 7.7 | N/A |

Table 11. Average hourly ozone loss rates in the Uinta Basin [34].

| Level (m) | 31 January–1 February | 1 February–2 February | 2 February–3 February | 3 February–4 February | 4 February–5 February | 5 February–6 February |
|---|---|---|---|---|---|---|
| Ouray Loss Rates (ppbv/h) | | | | | | |
| 2 | 2.5 | 1.7 | 2.4 | 2.1 | 0.9 | 1.6 |
| 50 | 2.5 | 1.8 | 2.2 | 2.1 | 0.8 | 1.6 |
| 100 | 2.3 | 2.0 | 1.6 | 1.8 | 0.9 | 1.4 |
| 150 | 2.0 | 1.8 | 1.3 | 1.4 | 0.8 | 1.2 |
| Fantasy Canyon Loss Rates (ppbv/h) | | | | | | |
| 2 | 2.4 | 2.6 | 2.1 | 2.3 | 1.8 | 1.8 |
| 50 | 2.3 | 2.1 | 1.5 | 2.7 | 1.6 | 1.8 |
| 100 | 2.4 | 2.0 | 1.0 | 1.9 | 1.7 | 1.2 |
| 150 | 1.7 | 2.3 | 1.2 | 2.8 | 2.0 | 1.8 |

Matichuk et al. applied the WRF and CMAQ models to the previously discussed February 2013 inversion [35]. It was found that the standard CMAQ model did not perform well. The WRF model performed better than the CMAQ simulations. This led to suggestions for improvements to the CMAQ model that would lead to better VOC, $NO_x$, and $O_3$ simulations.

Mansfield et al. conducted a statistical analysis of ozone in the Uinta Basin using two models: Uinta-5 and Uinta-8 [36]. Uinta-8 proved more accurate than Uinta-5 with a standard error of 11 ppb and 90% accuracy. He introduced a pseudo-lapse rate into their model as a way of determining inversion intensity.

There has been a measured decrease in ozone and $NO_x$ concentrations in the Uinta Basin, with a decrease of 3 ppb/year for ozone and 0.3 ppb/year for $NO_x$ [37]. Ozone concentrations are approximately 67% of what they were a decade ago, and $NO_x$ has declined to 58% of its concentration ten years ago. Organics have dropped to 40% of their concentrations measured in 2017. Mansfield et al. attribute these decreases to a decrease in global demand for fossil fuels leading to less production and to pollution controls, both government-mandated and voluntary, decreasing targeted emissions.

*6.2. Volatile Organic Compound (VOC) Studies*

An investigation into methane emissions in the Uinta Basin showed that $55 \pm 15 \times 10^3$ kgh$^{-1}$ was emitted in one day, 3 February 2012 [38]. Karion et al. concluded that this methane leak rate from gas and oil operations in the Uinta Basin, 6.2–11.7%, negates short-term climate benefits, defined as <70 years, of using natural gas over coal or oil. This conclusion was defined by the effect of methane on global warming. It was also found that the measurements that Karion et al. made exceeded the previous bottom-up estimates of the leak rate by a factor of 1.8.

A study that included the winters of 2012 and 2013 was conducted by Helmig et al. to measure the levels of nonmethane hydrocarbons (NMHC) in the Uinta Basin [39]. The annual flux of $C_2$–$C_7$ NMHC was found to be $194 \pm 56 \times 10^6$ kg year$^{-1}$. Benzene emissions were estimated to be $1.6 \pm 0.4 \times 10^6$ kg year$^{-1}$ and toluene emissions $2.0 \pm 0.5 \times 10^6$ kg year$^{-1}$. These estimates were used to identify a correlation between oil and gas emissions, increases in air toxics, and ozone production. Their paper includes speciated NMHC concentrations measured in the Uinta Basin.

Lyman et al. measured hydrocarbon, alcohol, and $CO_2$ fluxes from the Uinta Basin water ponds from 2013–2016 [40]. These ponds are from water that is used in the mining of oil and gas and are used as storage and disposal ponds. Production water comes from a well, along with oil or gas, and is disposed of by being pumped into these ponds to be evaporated. Inverse modeling methods were used to estimate emissions. It was found that alcohols accounted for 34% of the organic compounds and were the most abundant organic compounds (91% of VOCs). NMHC was found to be most of the organics emitted from water. Fluxes of formaldehyde were low (1–2%). It was found that the number and intensity of fluxes were affected by the time the water was exposed to the atmosphere and that ice impacted the fluxes. Part two of this study was completed by Tran et al. [41]. It was found that inverse modeling yielded inflated concentrations, with the largest discrepancy being in alcohols. This was used in the WATER9 wastewater emission model, and it was found that the flux chamber may be underestimating emissions. This may be from the inability of the models to adequately represent pond area, and the difficulty in modeling the ponds was most pronounced with some pond geometries and facilities. However, both models found that ponds' emissions contributed to the basin's organic emissions. Part three of this study was completed by Mansfield et. al [42]. This study attempted to quantify emissions from ponds by finding correlations between the concentration of volatile compounds in the ponds and their fluxes into the atmosphere. Emissions were generally found to satisfy the mass-transfer law, or Henry's Law, where the mass-transfer coefficient, or transfer velocity, is a function of wind speed, water and air temperature, and the partition coefficient between the air and water phases. These values were determined for a suite of emissions and compared with WATER9's semi-empirical algorithm. Predictions of WATER9 were found to be biased high but within an order of magnitude. Predications by the mass-transfer laws found that the concentration of any compound in water is proportional to its flux and also to within an order of magnitude. The authors conclude that the data suggests that hydrocarbons partition between aqueous solution and suspension. The salinity of the ponds was not found to correlate with mass-transfer coefficients. This led the authors to suggest that hydrocarbon transfer is water-film controlled. Alcohols are hypothesized to have larger mass-transfer coefficients in saline waters, but the authors were unable to confirm this. $CO_2$ emissions are hypothesized to be a result of the oxidation of hydrocarbons as older waters were found to have a higher portion of their emissions being $CO_2$. Under the assumption of their results scaling, the authors estimate the emissions from ponds in the Uinta Basin to be 700, 18,000, 1500, 700, and 5200 t/year of methane, $CO_2$, alkanes, aromatics, and alcohols, respectively. The total non-methane organic emissions are estimated to be 7500 t/year.

Ground-based measurements and the application of an atmospheric transport model were applied to the Uinta Basin by Foster et al. in order to quantify the amount of methane leaking from the oil and gas industry in the region [43]. It was found that the model worked well and that uncertainty due to meteorological conditions would not change the study's conclusions. It was found that estimates of a methane leak rate of $55 \times 10^3$ kg h$^{-1}$ by Karion et al., were correct [29]. The National Oceanic and Atmospheric Administration (NOAA) model disagreed with the EPA's estimates of methane leakage in the region but agreed with Karion's measurements. Both the NOAA and EPA simulations, however, underestimate nighttime and morning concentrations of $CH_4$ in some regions showing the uncertainty in spatial allocation. The NOAA model does, however, show that estimating methane emission rates was good enough to achieve promising results from a single dataset.

Foster et al., created a Basin-constrained Emissions Estimate (BEE) model to better understand methane emission rates in the Unita Basin [44]. The simulations yielded emissions between $44.60 \pm 9.66 \times 10^3$ and $61.82 \pm 19.76 \times 10^3$ kg $CH_4$ hr$^{-1}$. There are two major advantages to this model: 1. This model allows for simulations to be run more frequently, leading to the need for fewer aircraft campaigns, 2. The model can integrate daytime and nighttime emissions of $CH_4$ into the volume of a cold pool.

Lyman et al. deployed an infrared optical gas imaging helicopter to image 3428 oil and gas facilities in the Uinta Basin in early 2018 [45]. Lyman and his group also ground-surveyed 419 well pads. Despite the colder air temperatures impairing the detection of the plumes, many plumes were found. The predominant wells with plumes were newer, more productive, mostly oil well pads, and had a higher ratio of liquid storage tanks than other well pads. The majority of the plumes, 75.9%, were from liquid storage tanks, pressure release valves, piping, and thief hatches on the tanks. It was found that the storage tanks with control devices, such as combustors or vapor recovery units, were leaking before these control devices.

Spatial trends for many emissions were found in the Uinta Basin during a study by Lyman et al. during the winter of 2019/2020 [46]. Alkenes and acetylene were found to be more abundant in oil-producing regions of the basin and were calculated to comprise 25% of the total ozone formation potential from organics in these regions. Some of these hydrocarbons are from combustion, but Lyman et al. theorize that natural gas-fueled engines may be significant contributors. Total alkene and acetylene concentrations were 267 µg/m$^3$ in 340+ wells within 10 km and 12 µg/m$^3$ in regions with 15 or fewer wells.

*6.3. Combination Studies*

$CO_2$ and methane sensor gas networks have been implemented in the Uinta Basin based on previously made networks [47]. Instrumentation was updated, protocols standardized, and a new interpolation method was developed to allow these networks to take remote measurements with less maintenance and the need for frequent calibration. The new method for estimating uncertainty can be applied to other trace gas datasets with calibration information. The low uncertainty, 0.38 ppm for $CO_2$ and 2.8 ppb for methane shows promise for further application of these networks in tracking regional emissions.

Prenni et al. investigated aerosol and quantified haze at the Dinosaur National Monument on the northeastern rim of the Uinta Basin [48]. The study went from November 2018 to May 2020 during which ozone, speciated fine and coarse aerosols, inorganic gasses, and VOCs were monitored. During the first winter, three NAAQS exceedances of ozone were observed but no exceedances were observed during the second winter. High particulate concentrations were observed during both winters with 24-h averaged particle light extinctions exceeding 100 Mm$^{-1}$. Ammonium nitrate was the dominant species in the particulates and correlated well with particulate organics. It was found that nitric acid was the limiting reagent in ammonium nitrate formation. Prenni et al. recommend that a decrease in regional $NO_x$ emissions would lead to better visibility at the monument. VOC contributions, which were higher during the winter, were found to be from local oil and gas activity. Particulate concentrations during the summer months were far lower and the composition was found to shift from ammonium nitrate to coarse mass and soil.

Pollutants from 58 natural gas-fueled pumpjack engines in the Uinta Basin were analyzed from January 2021 to May 2021 [49]. It was found that the air-fuel equivalence ratio, the ratio between the air taken into the engine and the air needed for the combustion of fuel, could be used as a predictor of emissions. High air-fuel equivalence ratios lead to low $NO_x$ emissions and high VOC emissions while low air-fuel equivalence ratios were indicative of high $NO_x$ emissions and lower VOC emissions. 34% of engines had an air-fuel equivalence ratio of 3 which saw 57% of the fuel passing through the engine not combusted and exhaust gas containing a median of 3 ppm $NO_x$. Average $NO_x$ emissions were found to be only 9% of emissions in the regulatory oil and gas inventory while VOCs were found to be fifteen times higher than in the inventory. Lyman et al. hypothesized that this could be due to engines being aged and operating at lower loads.

*6.4. Uinta Basin Summary*

A substantial amount of modeling has been executed to better understand the photochemistry occurring in the Uinta Basin that leads to ozone formation. The focus of studies in the Uinta Basin has been ozone which is affected by VOC and $NO_x$ levels. Ozone, $NO_x$, and VOCs such as methane and other hydrocarbons/organics have been monitored. Many NAAQS pollutants have been neglected from study including CO, Pb, $NO_x$, and $SO_2$. There is a lack of peer-reviewed PM investigations and source apportionment studies in this region. While this region does not currently suffer from PM exceedances, baseline measurements conducted now would be helpful in identifying trends in the NAAQS standards occurring in the future.

The wintertime ozone exceedances in the Uinta Basin are of interest and have been an anomaly worth studying. The Uinta Basin is full of ozone precursors due to anthropogenic emissions from the harvesting of fossil fuels. This combined with the increased surface albedo from snow cover has led to wintertime ozone spikes. This surface albedo may be further enhanced as ozone precursors and other species, such as methanol and formaldehyde, that are found in the basin from anthropogenic activities add to the composition of the snow. This abnormal phenomenon may continue until the introduction of VOC and $NO_x$ precursors to ozone are better controlled or until snow is no longer present in the basin. This is evidenced by studies of ozone during years when there was less snow cover showing fewer ozone spikes.

## 7. Salt Lake Valley

The air quality in this region has been described in thirty-four publications between 2002 and 2022. The Utah DEQ has fourteen active air monitoring sites in the Salt Lake Valley Region (Harrisville—490571003, Bountiful—490110004, Antelope Island—490116001, Hawthorne—490353006, Herriman—490353013, Copperview—490352005, Inland Port—490353016, Lake Park—490353014, Near Road—490354002, Rose Park—490353010, Saltair—490353005, Utah Technical Center—490353015, Erda—490450004, Badger Island—490456001) [15]. This is a highly populated urban area situated between two mountain ranges in a valley. This makes Salt Lake Valley prone to persistent cold-air pool events which lead to elevated levels of $PM_{2.5}$. This valley also suffers from summertime ozone. It is also adjacent to the Great Salt Lake which is a source of halogens for the area. This makes the study of the lake breeze an interesting phenomenon.

### 7.1. Particulate Matter (PM) Studies

Peterson et al. used time-of-flight static secondary-ion mass spectrometry (TOF-SIMS) to analyze the surface organic layer of atmospheric aerosol particles [50]. Cascade impactors were used to collect dust from various places including during an Asian Dust Event in Salt Lake City in April 2001. The dust from this storm could be traced across the Pacific Ocean using NASA's Total Ozone Mapping Spectrometer. Filters that collected urban samples appeared black while samples containing Asian dust were brown. It was found that brown PM was mineral dust containing iron minerals such as hematite and goethite, which imparted a reddish color to the urban aerosols that appeared black. The positive TOF-SIMS spectra were dominated by sub-micrometer organics, while the negative spectra showed nitrates and sulfates. The Asian dust samples were found to be dominated by super-micrometer-sized mineral aerosol.

Silcox et al. investigated how $PM_{2.5}$ concentrations change during cold-air pools from 1 January to 20 February 2011 [51]. It was found that $PM_{2.5}$ concentrations rose as the valley heat deficit rose. This indicates that the longer the cold-air pool event, the higher the $PM_{2.5}$ concentrations will be as the heat deficit continues during cold-air pool events. Concentrations of $PM_{2.5}$ were highest on the valley floor and decreased with elevation. During the study, it was found that $PM_{2.5}$ concentrations can increase by 6 $\mu g/m^3$/day. Breakup events were seen to lower $PM_{2.5}$ concentrations at higher altitudes before clearing $PM_{2.5}$ near the valley floor.

Positive matrix factorization and the EPA's Unmix Model were used to identify factors contributing to $PM_{2.5}$ from data collected between 2007 and 2011 for three regions along the Wasatch Front: Lindon, Salt Lake City, and Bountiful [52]. During cold-pool events, both factor analysis techniques showed secondary inorganic aerosols to be the main contributors to $PM_{2.5}$ (60–80% in all locations). Secondary $PM_{2.5}$ was a larger portion of the $PM_{2.5}$ concentration when concentrations were higher than 20 $\mu g/m^3$. Secondary ammonium nitrate was determined to be the top contributor; however, secondary ammonium chloride (10–15% when 24-h averages of $PM_{2.5}$ exceeded 30 $\mu g/m^3$) was also identified as a factor. Ion balance analysis was used to confirm the presence of ammonium chloride. Sources of primary $PM_{2.5}$ formation were identified as wood smoke and gasoline emissions. Dust may be part of the $PM_{2.5}$ in the fall.

The composition and secondary formation of particulate matter in the Salt Lake Valley from January to February 2009 was studied by Kuprov et al. [53]. Hourly averages of $PM_{10-2.5}$, $PM_{2.5}$, $NO_x$, $NO_2$, NO, $O_3$, CO, $NH_3$, particulate-phase nitrate, nitrite, sulfate, gas-phase HONO, $HNO_3$, and $SO_2$ were measured. In the absence of inversions, ammonium nitrate averaged 40% of the total $PM_{2.5}$ concentrations as opposed to 69% during inversions. Nitric acid was the limiting agent in ammonium nitrate formation, leaving large concentrations of excess ammonia. Seven percent of the $PM_{2.5}$ during inversions was composed of sulfates and nitrates as opposed to 11% during non-inversions periods. Ozone levels remained below EPA standards for the duration of the study, and the Salt Lake Valley was found to be oxidant and VOC deficient regarding ozone production. It was theorized that decreasing $NO_x$ concentrations during inversions would decrease ammonium nitrate formation, with the acknowledgment that this would increase ozone concentrations. Therefore, a complete ozone isopleth was recommended before implementing a $NO_x$ limiting policy.

In 2014 Whiteman et al. explored the relationship between PM and meteorological conditions in the Salt Lake Valley and found that there has been a decrease in PM in the valley over the previous 40 years [54]. It was found that daily average concentrations of $PM_{2.5}$ were usually exceeded during high atmospheric stability episodes where concentrations of $PM_{2.5}$ could rise by 10 $\mu g\ m^{-3}\ day^{-1}$ for 4–5 days, or about 60 $\mu g\ m^{-3}$, days before it plateaued. Exceedances were also more prevalent if snow cover was on the ground, when surface pressure and RH were high, and when temperatures and wind speeds were low. Solar reflection from the snow, high RH, and low temperatures all contribute to the deliquescence point of $PM_{2.5}$ and its secondary formation. Multiple linear regression was performed on meteorological variables to estimate daily $PM_{2.5}$ concentrations and yielded a correlation factor of 0.81.

The impacts of the 2007 and 2012 wildfires in the western United States on CO, $CO_2$, and $PM_{2.5}$ concentrations in the Salt Lake Valley were studied by Mallia et al. [55]. The Weather Research and Forecasting-Stochastic Time-Inverted Lagrangian Transport (WRF-STILT) model was used to resolve the growth/decay of the planetary boundary layer (PBL) and the winds used for backward trajectory analysis. Simulated values were reasonably comparable to observational values. The model showed that $CO_2$ contributions from the wildfires were negligible and difficult to deconvolve during the 2007 wildfires (2 ppm) compared to anthropogenic contributions (20–40 ppm) and model uncertainty. CO contributions were minimal, except for two days when they increased by a factor of 1.5 in 2007. CO concentrations in 2012 sometimes exceeded anthropogenic emissions. The uncertainty in CO concentrations from the model was calculated to be approximately 15 ppb. The wildfires' primary contributions to $PM_{2.5}$ were substantial. Throughout this study, it was observed that the fires contributed as much as 250 ppb of CO (3-h average), 15 $\mu g/m^3$ of $PM_{2.5}$ (24-h average), and 2 ppm of $CO_2$.

Brown et al. measured particle matter count and black carbon at Adcock Elementary, Las Vegas, NV (urban site next to a freeway) and Hunter High School, West Valley, UT (Urban site) [56]. Emissions from the freeway produced high particulate and BC concentrations at the Adcock site. At Hunter High School, winds did not appear to affect the BC or particulate concentrations analogous to what was observed at Adcock elementary. In the mornings, large spikes were seen at both schools due to commuter traffic. It was noted that traffic speed, volume, or fleet type did not affect Adcock's concentrations. Other peaks were seen at midday and in the afternoon and were ascribed to new particle formation under the favorable conditions of low wind speeds and high solar radiation.

A wintertime cold-air pool event in 2016 was studied to see how $PM_{2.5}$ and $N_2O_5$ were affected [57]. Some pollutants, $NO_x$, and CO, doubled during the event, but ozone decreased almost threefold. It was found that the atmosphere's composition during the event varies greatly with altitude at night, with lower $NO_x$ and higher ozone and $N_2O_5$. Evidence was found that $N_2O_5$ plays a role in nitrate formation. Chemistry was found in the upper levels of the cold-air pool that played a role in nitrate formation during the nighttime and early mornings that was then mixed during the daytime and enhanced $PM_{2.5}$ concentrations throughout the cold-air pool.

Black carbon (BC) and organic carbon (OC) were measured in 31 samples collected in three locations in Salt Lake City from 2012–2014 [58]. Analysis of the samples for $^{14}C$ showed that fossil fuels contributed 88% of BC and 58% of OC and were the dominant source of carbonaceous material in particulate matter during the winter. Source contributions of BC and OC did not change during inversion events leading to an increase in their concentrations around the city. BC remained constant in all seasons; however, OC increased to 62% of the biomass component. It was concluded that for the region to meet EPA air quality standards more stringent policies regarding the use of fossil fuels are needed.

Odd oxygen, or $O_{x,Total}$, which includes $O_3$ and particulate nitrate ($pNO_3^-$), was studied to determine that odd oxygen products originate from the same cycle and are determined by $NO_x$/VOC ratios [59]. This was executed to understand the response of ammonium nitrate aerosol production to emission controls in a $HNO_3$ sensitive environment. It was found that the Salt Lake Valley was well above the saturation limit for $NO_x$, so the focus was shifted to limiting VOCs as a control. The analysis found that a decrease in $PM_{2.5}$ in the area was due to a decrease in VOC emissions and that a decrease in $NO_x$ may slow the decrease in ammonium nitrate-dominated aerosol regions.

McDuffie et al. investigated the contributions to $PM_{2.5}$ formation from nocturnal heterogeneous reactive nitrogen during wintertime pollution events [60]. An observationally constrained chemical box model suggested that $PM_{2.5}$ was formed nocturnally during pollution events in the Salt Lake Valley. Production rates of nitrate had a median of 1.6 $\mu g/m^3$ h and the heterogenous uptake coefficient of $N_2O_5$ ($\gamma(N_2O_5)$), and the production yield of $ClNO_2$ ($\varphi(ClNO_2)$) had medians of 0.076 and 0.220. Modeled nighttime nitrate production levels (9.9 $\mu g/m^3$ night) combined with the $P_{NO_3^-}$ max values (10.6 $\mu g/m^3$ night) indicate that nitrate production is more sensitive to $NO_2$ than $\gamma(N_2O_5)$. The highest loss rate was through air parcel dilution (42%) with a dilution of $1.3 \times 10^{-5}$ $s^{-1}$. Effects of 24-h dilution predicted a reduced median of 2.4–3.9 $\mu g/m^3$ day of nitrate.

A source apportionment study executed at the Neil Armstrong Academy in January–February 2016 showed the sources of fine particulate matter material in the western region of the Salt Lake Valley [61]. A PMF analysis of hourly averaged data, which improved the PMF analysis over daily average data, yielded a tally of seven factors contributing to $PM_{2.5}$ formation. Automotive and diesel traffic, emission from the copper smelter, secondary formation of ammonium nitrate, and wood smoke fires incorporated the seven factors. Wood smoke was broken into three factors: primary wood smoke particulates and two secondary aerosol formation emission factors. The hourly average rate of $PM_{2.5}$ production during an extended cold-pool event was 25.2 $\mu g/m^3$ h. The source apportionment data are found below in Table 12.

Table 12. Source apportionment from the Neil Armstrong Academy study.

| Factor | Hourly Average $PM_{2.5}$ Concentration ($\mu g/m^3$) | Percentage (%) |
| --- | --- | --- |
| Diesel | 2.5 | 11.7 |
| Nitrate | 12.12 | 56.5 |
| Smelter | 2.27 | 12.7 |
| Organic Material | 1.11 | 5.2 |
| Automobiles | 2.56 | 12.0 |
| Wood Smoke | 0.39 | 1.8 |

Much of the aerosol in Utah is nitrate aerosol, and because of this, studies have been conducted to investigate the sources and distribution of nitrate compounds around the Great Salt Lake [62]. Particulate ammonium ($pNH_4$) and ammonia ($NH_3$) were sampled at sites around the Great Salt Lake, Northern Utah, and Utah Valley. The highest concentrations of $NH_x$ ($NH_3$ and $pNH_4$) were found in Cache Valley, attributed to the concentration of livestock, but were not much higher than in Salt Lake or Utah Valleys. The model used in this study, STILT, suggested that UDAQ estimates livestock emissions are underestimated by up to a factor of 4.5 during some periods. It is suggested that facility-based measurements for livestock emissions be made in the future. It was found that two-thirds of $NH_x$ in the Salt Lake Valley originated from within the valley. During cold-pool air events, the other third is carried into the Salt Lake Valley, mostly from the Utah Valley. 70% of $NO_x$ leading to $PM_{2.5}$ in Cache Valley was found to be transported from other counties into Cache Valley.

In 2020, a high-density network of low-cost sensors was deployed in Salt Lake City to monitor smoke from nearby wildfires [63]. The low-cost sensors were mounted onto the roof of the light-rail public transportation system in the Salt Lake City area. The data collected were also used to evaluate a coupled fire-atmosphere model. The root-mean-square value of 8–14 µg m$^{-3}$, a 10% bias, suggested that the low-cost sensors can effectively measure smoke plumes and characterize the spatiotemporal heterogeneity of the smoke plumes. The mobile measurement platform also provided insights into smoke simulations and high-resolution fire-atmosphere modeling.

Community-based measurements were made by deploying a low-cost network of $PM_{2.5}$ sensors in the Salt Lake Valley during the 4 July 2018, fireworks, the 5 and 6 July 2018, wildfire, and the 3–7 January persistent cold-air pool [64]. A Gaussian Process Model was used to predict the reference measurements during all three periods. Table 13 shows the results of the model:

Table 13. Prediction statistics from the Gaussian Process Model.

| Event | N | Hourly Root-Mean Squared Error (µg/m$^3$) | Normalized RMSE (%) |
| --- | --- | --- | --- |
| Fireworks | 16 | 12.3–21.5 | 14.9–24 |
| Wildfire | 46 | 2.6–4.0 | 13.1–22.9 |
| Cold-Air Pool | 96 | 4.9–5.7 | 20.2–21.3 |

Geospatial differences in $PM_{2.5}$ were found that were not highlighted by government measurements. This shows the potential for low-cost sensor networks, coupled with data-fusion algorithms, in dynamically estimating $PM_{2.5}$ concentrations.

Ambient ion monitoring chromatography was used to find gas-phase $NH_3$, $HNO_3$, HCl, $SO_2$, $NH_4^+$, $Na^+$, $K^+$, $Mg^{2+}$, $Ca^{2+}$, $NO_3^-$, $Cl^-$, and $SO_4^{2-}$ between 21 January and 21 February 2017 [65]. During this persistent cold-air pool event, secondary $NH_4NO_3$ was the dominant component of particulate matter in the Salt Lake Valley. Non-volatile cation ions were found in the measured $PM_{2.5}$ (p$Na^+$, p$K^+$, p$Ca^+$, p$Mg^+$), suggesting mineral salt or dust is present during these cold-air events. These could produce a sink for $HNO_3$, which could be taken up into NaCl or $CaCO_3$-containing particles. This coarse form of nitrate can be supported by the correlation between $Ca^{2+}$ and $NO_3^-$ in the snowpack where $NO_3^-$ is 2.8 times larger than $NH_4^+$, which suggests that some $NO_3^-$ is not found in the $PM_{2.5}$ or gas-phase nitrates.

The effects on air quality of two anti-idling campaigns between September 2019 and February 2020 [66]. A 38% decrease in idling time and an 11% decrease in the number of vehicles idling at school drop-off zones were observed. An improvement in air quality was noticed in the middle of the campaign, however, substantial effects on ambient pollution concentrations were caused by seasonal variability and inversion events.

*7.2. Dust Studies*

Dust events from 1930–2010 were studied by Steenburgh et al. and were found to have considerable inter-annual variability [67]. These dust events have contributed to dust loading in the snowpack. There has been a decline in dust-event frequency during the study period which peaked in spring and autumn. The Sevier Desert, Sevier Dry Lakebed, Escalante Desert, Milford Valley, and the West Desert of Utah are found to contribute to these dust storms.

Two dust event days on 19 April 2008 and 4 May 2009, were investigated for their role in producing elevated $PM_{10}$ levels in Salt Lake City [68]. Cyclonic systems that originate in Nevada are the primary producers of dust storms in the Great Basin. They produce southwesterly winds that lift sediment into the atmosphere from a known hotspot: Sevier Dry Lake, Tule Dry Lake, the Great Salt Lake, and Milford Flats. These regions are "hot spots" because they contain deflatable dust-sized particles, low friction velocities at surface levels enhanced by low moisture content in the soil, and sufficient wind speeds to loft up sediment. The mountain-valley topography funnels the winds that carry the sediment to Salt Lake City. Most of these dust event days happen in the springtime, and sixteen of these dust event days have been recorded over the eighteen years prior to this study.

Sixteen dust days between 2004 and 2010 were used to identify seven primary and five secondary areas responsible for dust plumes in the eastern part of the Great Basin which extends from Nevada into Utah [69]. Barren land cover was responsible for 60% of the observed dust plumes, and 75% of the dust from vegetated land cover areas originated from the recently burnt Milford Flats. Many of the dust source regions were found to be regions that correspond to the remnants of Bonneville Lake, a Late Pleistocene paleolake that stretched from the Sevier Basin to the Great Salt Lake Basin in Utah. Some areas, such as parts of the Great Salt Lake Desert and Sevier Dry Lake, are influenced by military and agricultural activity. They concluded that dust plumes are evidence of "hot spots" in the Great Basin that generate dust and that the Basin is not a uniform emitter of dust.

A dust storm from 14–15 April 2015 was analyzed by Nicoll et al. to document its development and impact [70]. Dust was deposited on the snow in the Wasatch Mountain Range near Salt Lake City. This was the first study to analyze dust deposits in a snowpack from a single dust storm that was analyzed for dust mobilization, air quality changes, and dust deposition rather than seasonally deposited dust. Gusts in the frontal system were measured between 60 and 80 mph. The winds caused erosion of silt- and clay-rich areas of the western desert region. The Sevier Dry Lake caused a deflation which resulted in a dust plume that stretched 155 miles to the north. The dust layer from this single storm was found to be 1–3 cm thick and contain a medium dust particle size of 10.81–12.55 μm in diameter. The radiative properties of this dust layer were similar to previously measured aggregated dust from this region. This layer was enriched in As, Cd, Cu, and Mo by up to 10× relative to the average elemental concentrations in the earth's upper continental crust. The heavy metals found in the dust, Cu, Pb, As, Cd, Mo, and Zn, were thought to come from mining activities in the region and anthropogenic activities near Las Vegas, NV, and smog plumes in California.

Putman et al. analyzed the dust deposited in 18 passive dust samplers, positioned near the Great Salt Lake, for two sampling periods that ran from Fall 2018–Spring 2019 and Spring 2019–Fall 2019 [71]. The highest dust fluxes were found to be at the sites near the Great Salt Lake (GSL) playa. The $^{87}Sr/^{86}Sr$ ratio was used to determine that the dust regularly measured in the urban corridor is due to local soil materials. Metals in the dust particles were attributed to metal-emitting industrial activities such as mining, oil refining, pesticides, and herbicides. Lanthanum was found to originate at oil refineries with possible transport from the GSL playa. Sources of thallium include mining, smelting, metallurgy, fossil fuel combustion, the electronics industry, and waste incineration. Copper and molybdenum were associated with mining activities. Arsenic and vanadium were in lower abundance at the sites near the GSL playa and higher at southern sites far from the GSL playa. Putman et al. concluded that arsenic and vanadium are linked and unlikely to arise from the playa but can be traced to historic herbicide and pesticide application to orchards and crops in the early 1900s. Another possible source is that of the Kennecott mine or smelting activities. The authors recognize that they may not have the spatial or temporal granularity to tease apart contributions of each of these sources and recognize possible contributions from the GSL.

*7.3. Ozone Studies*

In 2009, the State of Utah introduced a program named the Clean Air Act Challenge to educate the population on actions that could be taken to reduce ozone and PM concentrations. The challenge was undertaken annually and lasted for a month. The program recommended actions such as participating in carpools, using mass transit, using active transportation, teleworking, skipping trips, and chaining trips together. The Clear the Air Challenge (CAC) program's effectiveness in the Salt Lake Valley was evaluated for its ability to reduce ground-level ozone during the summer [72]. A range of nonequivalent control group models was created, and only one model showed statistically significant reductions in ground-level ozone (3.6%). It was concluded that the implementation of the program did not yield meaningful changes in air quality sufficient to affect public health and that voluntary programs may not be effective in improving air quality in Utah.

The 2015 Great Salt Lake Summer Ozone Study (GSLSO3S) involved twenty-four fixed sampling sites and several mobile sampling sites on vehicles, light rail, and helicopters [73]. This study was designed to measure the spatial and temporal variations in ozone near the Great Salt Lake. The pattern of diurnal cycles of ozone was seen in urban areas with low concentrations in the mornings and at night, contrasting with higher concentrations during the day. Concentrations of ozone in rural sites seemed to be dictated by the wind and valley elevations more than the nighttime ozone titration. The highest ozone concentrations were found in the urban areas southeast of the Great Salt Lake. In June 2015, measured high ozone concentrations resulted from local primary processes. This changed in August 2015 when ozone concentrations were found to be impacted by smoke from wildfires being carried into the area by wind circulation.

In mid-June 2015, the influence of lake breezes from the Great Salt Lake on ozone concentrations in the region was investigated [74]. The highest midday ozone concentrations were found at Farmington Bay. On 17 June, the lake breeze was found to transport lower ozone concentrations from the northwest part of the Great Salt Lake towards the southeast region. The boundary layer in the breeze carried a gradient of almost 20 ppb difference between the leading edge and the wake of the breeze.

The Utah Transit Authority light-rail Transit Express (TRAX) system is a transportation system for light-rail electric cars in the Salt Lake Valley area. Two cars from this system were fitted with ozone and $PM_{2.5}$ sensors to measure air quality in the region [75]. This study presented a high spatial and temporal resolution gradient of ozone and $PM_{2.5}$ in this region. Pollutant hotspots were able to be identified during pollution events. It was found that ozone concentrations were higher in parts of the valley that were higher in elevation. Ozone levels were found to be lower near the roads, and it was theorized that this is because of $NO_x$ titration from the $NO_x$ produced from vehicle emissions. While these pollution gradients are helpful, it was noted that the observations represent temporal resolution at locations while the sensor moves around on the train. An example of this was that the train was far away from the stadium where fireworks were being set off and did not register the air quality in that region at that time.

The electric-powered light rail transit system (TRAX) in the Salt Lake Valley in Utah was used as a means for mobile measurements of $CO_2$, $CH_4$, $O_3$, and $PM_{2.5}$ throughout the Salt Lake Valley from December 2014 to April 2017 [76]. It was noted that all measurements are surface measurements and cannot characterize vertical distributions of emissions in the region. $CO_2$ was higher in the northern metropolitan area and decreased in the southwestern area bordering suburban areas. $CH_4$ existed indistinct plumes throughout the valley in contrast to the broad $CO_2$ features. The plumes were near industrial sources such as natural gas power plants and a brick factory that uses natural gas to power its furnace. These plumes were observed throughout the day, but one plume was seen only during the daytime hour, indicating commercial/manufacturing activity. These data showed that mobile measurements with few passes through an area or only at specific times might miss sources of emissions that vary during the day or from day to day. A plume (20 µg/m³) of $PM_{2.5}$ was found near the gravel pit in the south-central region of the valley. After a pollution episode, the gradient shifted, and higher $PM_{2.5}$ concentrations (20–30 µg/m³) could be found in the southern parts of the valley. The summer average of $PM_{2.5}$ was below the NAAQS standard. Ozone was found to be 5–10 ppb lower in the urban regions of the valley than in the suburban foothills. A strong anti-correlation with $NO_2$ was found, and $NO_2$ patterns seemed to correlate with $CO_2$ patterns of broad gradients across the valley. The largest $NO_2$ plume was seen at the locomotive rail yard.

*7.4. Volatile Organic Compound (VOC) Studies*

Sources of formaldehyde in Bountiful, UT, were investigated by Bhardwaj et al. [77]. PMF analysis showed that primary sources of formaldehyde are associated with biomass burning and the photooxidation of biogenic emissions. This accounted for 79% of the formaldehyde, while the other 21% can be accounted for by anthropogenic emissions such as mobile emissions and industrial and refinery emissions. It was found that these emissions are probably mostly from the oil refineries as they are an order of magnitude larger than other industrial sources. The observed diel pattern suggested a coupling of formaldehyde concentration with actinic flux showing that VOC photooxidation plays an important role in formaldehyde formation. The conversion of benzene, toluene, ethylbenzene, and xylenes (BTEX) into formaldehyde was slow, and inconsistency was found in the appearance of their peaks; thus, $O_3$ must have been involved in the oxidation processes in the atmosphere. Back-trajectory wind analysis showed that the largest sources of formaldehyde were in the southwest, where the wind would blow over the oil refineries in the North Salt Lake area.

*7.5. Other Pollutant Studies*

The relationship between $CO_2$ as a trace gas and other pollutants was studied by Bares et al. during the winter of 2015–2016 [78]. A correlation between $CO_2$ and $PM_{2.5}$ was found across all seasons; however, the relationship changes during times of inversion when secondary aerosol formation is a major contributor to $PM_{2.5}$ concentrations. A covariance was found between $CO_2$ and $PM_{2.5}$ during inversion events which can be used to estimate $PM_{2.5}$ concentrations. Baseline $CO_2$ ($CO_{2ex}$) correlations with CO, $CH_4$, and $NO_x$ during all winter conditions were also identified. $CO_{2ex}$ also provided insights into transportation versus the secondary formation of $PM_{2.5}$ during inversion events. It was found that current emission inventories overestimate CO and $CO_2$ by a factor of 1.3 and the $NO_x$: $CO_2$ ratio by a factor of 3.0.

*7.6. Meteorological Contributions to Air Quality Studies*

Bailey et al. published a paper in 2011 that described the use of the Global Telecommunications System (GTS) to characterize the relationship between changing temperatures and large-scale atmospheric circulation from 1994–2008 [79]. Six locations were investigated during this study, including Salt Lake City. The GTS data showed a decline in near-surface inversions in Salt Lake City. A strong link was found between large-scale atmospheric circulation and local inversion activity. It was theorized that Salt Lake Valley's topography helped facilitate inversions by creating stagnant conditions.

Ion concentrations in snowpacks along the Wasatch Mountains were measured by Hall et al. in 2011 [80]. Higher than average $Na^+$, $Cl^-$, $NH_4^+$, and $NO_3^-$ were found with concentrations of 120, 117, 42, and 39 µeq $l^{-1}$, respectively. Surface snow ion concentrations were highest at levels of 100–300 m above the valley floor and reached concentrations of 2500, 3600, 93, and 90 µeq $l^{-1}$ for $Na^+$, $Cl^-$, $NH_4^+$, and $NO_3^-$. Nitrogen deposition in the snow was measured to be 0.8 kg $ha^{-1}$.

Lake breezes from the Great Salt Lake during cold-air pool events were seen to be more complex to model than breezes during the summer but both were found to contribute to pollution in the Salt Lake Valley [81]. During the afternoons, there is evidence that these breezes introduce fine particulate matter while decreasing the vertical mixing depth of the cold-air pool and increasing its vertical stability.

Foster et al. conducted simulations to determine land use and snow cover effects on cold-air pool events in an effort to improve the WRF model [82]. Conclusions of this study show that cold-air pools are sensitive to surface land use and snow cover that improving the specification of land use and snow cover results in substantial changes in albedo, heat flux, temperatures, and boundary-layer cover, and that land use and snow cover affect the thermally-driven flows in the Salt Lake Valley.

Turbulence during cold-air pool events in the Salt Lake Valley showed that these events are associated with low surface temperatures and low wind speeds during the 2010/2011 winter [83]. Snow cover was found to be offset by strong winds through the valley and produced mixing in the boundary layer. Surface sensible (H) and latent (LE) heat fluxes were shown to be lower during strong prolonged cold-air pool events, which also correlated with a higher ratio of heat flux (H/Rn and LE/Rn) which was theorized to come from boundary layer clouds and could promote mixing. The largest turbulent fluxes were in the range of $-0.05 < \xi \geq -0.02$. The importance of dynamic land-use information in models was shown by a smaller surface exchange coefficient measured at the bare land site than at the short vegetation site. Part two of this study quantifies the ability of the WRF model to simulate land-atmosphere interactions during persistent cold-air pool events [84].

*7.7. Salt Lake Summary*

The Salt Lake Valley has been extensively studied. Elemental analysis of dust collected in Salt Lake City would be a strong contribution to the dust studies completed in this region. While this is the most densely populated region of Utah, it may be of interest to the rest of the state to shift some of the focus from this region to better study air quality in neighboring regions that may be impacted by emissions from the Salt Lake Valley. This region is currently in nonattainment for $PM_{2.5}$ but was found to be nitric acid limited for PM formation [57]. This region is currently a nonattainment region for $O_3$. While a substantial amount of research has been carried out to measure the spatial and temporal variation of $O_3$ in the valley, uncertainties remain as to the policies that may best curtail the high concentrations of ozone measured during the summer months. Measurements of speciated VOCs combined with the generation of ozone isopleths would contribute to crafting effective policies for $O_3$ reduction. This would assist in determining whether Salt Lake Valley is a $NO_x$-limited area or a $NO_x$-saturated area, as suggested by Womack et al. [71]. Contributing to the region's issues with high summertime $O_3$ concentrations is the influence of stratospheric incursion as well as the transport of both ozone precursors and ozone from regions as far west as Asia. Deconvolving these influences from local production of $O_3$ will be important in aiding the region in becoming compliant with the EPA's NAAQS.

## 8. General Conclusions and Recommendations

Between 2002 and 2022 there have been 68 published investigations on the air quality in Utah. The efforts in the regions of Utah Valley, Summit County, Southern Utah, Cache Valley, Uinta Basin, and Salt Lake Valley have been diverse. The research conducted in each region has not been divided equally based on population or deficiencies in data but instead based on interest and resources within the area. While other factors such as non-attainment classifications or rare or unusual air quality phenomena, such as wintertime ozone in the Uinta Basin, may be expected to skew research efforts towards specific areas, these factors have been seen to take precedence over population-driven research. Table 14 shows the number of papers published in each region and the percentage of the total efforts afforded to each region. As seen from these data, efforts do not follow population trends though a pattern can be identified that shows that the most studied regions are areas to which university researchers have easier access. While the Salt Lake Valley is the most populated region in Utah and is classified as a non-attainment area, it has been disproportionally studied when compared to other regions in the state. Utah Valley is the most understudied region when comparing the effort of research to population despite it also being a non-attainment area. Southern Utah has also been neglected when considering this ratio. These are also the two fastest-growing regions in Utah which warrants conducting air quality studies in these regions as a priority. The Uinta Basin is the lowest populated region but is second in research output due to its unique air quality phenomenon of wintertime ozone.

It was found that Pb, $SO_2$, and TSP have been largely neglected across the state. Measurement of TSP sometimes has been included in $PM_{2.5}$ studies, but Pb and $SO_2$ have not been monitored many times in the last 20 years. This may be because the ambient concentration of Pb and $SO_2$ have been far below the EPA standards in the past, and consequently do not merit continual monitoring.

**Table 14.** Research efforts for air quality in each region were evaluated by the number of published research articles per region between 2002–2022.

| Region | Number of Research Articles | Effort (%) | Population Ranking |
|---|---|---|---|
| Utah Valley | 3 | 4.4 | 2 |
| Summit County | 1 | 1.5 | 5 |
| Southern Utah | 3 | 4.4 | 3 |
| Cache Valley | 6 | 8.8 | 4 |
| Uinta Basin | 21 | 30.9 | 6 |
| Salt Lake Valley | 34 | 50 | 1 |

Table 14 shows the components of air quality that have been studied and the gaps in air quality monitoring by region. Table 14 shows that even though the total number of studies completed in a region does not match population numbers, the studies that have been executed in lesser-studied areas provide more complete measurements of the air quality in that region.

## 9. Addressing Non-attainment $PM_{2.5}$ Regions

A general lack of PM source apportionment studies conducted throughout the state of Utah is noted; with only two regions, Salt Lake, and Utah Valley regions, ever conducting this type of analysis. PM source apportionment studies can be an important tool in formulating effective policies for pollution reduction in each region. The most informative source apportionment studies involve hourly averaged measurements of a wide range of pollutants and precursor pollutants that contribute to the formation of $PM_{2.5}$ including all the compounds included in the NAAQS, speciated VOCs, BC, OC, and metrological data. The justification for this recommendation comes from observing the completeness of the dataset used in the source apportionment analysis conducted in Utah Valley and the conclusions that were discussed. Source apportionment studies may be rigorous but encompass many of the neglected areas of study, outlined in Table 15, in one campaign which is a conservative use of resources. Speciated VOC measurements could be executed using a gas-chromatography mass spectrometer (GC-MS) and should be operated continuously over long enough periods of time to capture seasonal changes. These data could be used in the source apportionment studies and to determine VOC trends. Knowing the VOC trends of the area would allow for the identification of the origin of VOCs that contribute to $PM_{2.5}$ formation. These VOC measurements could be used as input data for models that could determine the influence of secondary organic aerosol production on a region and the lifetime and ozone production potential of each of the measured VOCs. Speciated VOC measurements, along with measurements of wind speed and wind direction, can be used in source apportionment studies of VOCs. A source apportionment study in each of the major areas along the Wasatch Front can determine whether the major emitters of VOCs are anthropogenic emitters or biogenic emitters. If substantial anthropogenic emitters are identified, they can be specifically targeted for mitigation policies to not only reduce $PM_{2.5}$ production but to reduce ozone production in the process.

**Table 15.** Summary table of what has been studied in the literature in each of the defined regions of Utah.

| Region | Total Suspended Particles | Organic Material (VOCs) | Organic Material (Non-VOCs) | CO | Pb | Dust | NO$_x$ | O$_3$ | PM$_{2.5}$ | PM$_{10}$ | SO$_2$ | PMF Analysis | Black Carbon | Elemental Carbon | Ion Sources |
|---|---|---|---|---|---|---|---|---|---|---|---|---|---|---|---|
| Utah Valley | | [6,16,17,24] | [6,16,17] | [17] | | | [6,17] | [17] | [6,16] | [6,16] | | [6] | [6] | [6] | sulfate [16,24] nitrate [16,24] chloride [16,24] ammonium [16,24] sodium [6] |
| Summit County | | | | | | | | | [18] | | | | | | |
| Southern Utah | [20] | [20] | [20] | | | [19,20] | | | [20] | [20] | | | | | nitrate [21] ammonium [21] |
| Cache Valley | | [24,25] | [25] | | | | [21,25] | [24,25] | [4,5,21,23–25] | [24,25] | | | [4] | [4] | Sulfate [4,24] nitrate [4,21,24] chloride [4,24] ammonium [4,24,36] |
| Uinta Basin | | [27,33,37–46,49] | | | | | [27,32,35,37,49] | [27–37,48] | [37,48] | [48] | | | | | nitrate [48] ammonium [48] |
| Salt Lake Valley | | [24,50,61,76,77] | [50,62,76,77] | [53,55,57,78] | | [50,67–71] | [53,57,59,60,62,66,75,76,78] | [24,53,57,59,66,72–77] | [24,50–66,72,75,76,78,79] | [24,50,53,54,68] | [53] | [52,61,77] | [56,58] | [58] | sulfate [24,50,53,65] nitrate [24,50,53,59,61,62,65,80] chloride [24,53,65,80] ammonium [24,53,62,80] [65] sodium [65,80] potassium [65] magnesium [65] calcium [65] |

## 10. Addressing Non-attainment $O_3$ Regions

Table 1 shows that some regions in Utah struggle with nonattainment for $O_3$ as of the year 2022. It would be informative to study ozone on a regional and seasonal basis to investigate the regional versus local source contributions to ozone concentrations. These investigations would help to elucidate the baseline concentrations of ozone in Utah due to the regional transport of ozone into Utah compared to the contributions due to local sources. Additionally, studies should be completed to study the contribution of stratospheric ozone incursion events on the concentrations of ozone in Utah. It should be noted that work has been completed on understanding the regional transportation of ozone into Utah even if these reports do not appear in the peer-reviewed and published literature [85–87].

A strategy for ozone mitigation is to produce regional ozone isopleths, often on a seasonal basis. The major precursors to ozone formation are $NO_x$ and VOCs. Therefore, mitigation strategies for $O_3$ will generally involve the mitigation of $NO_x$ and VOC emissions. The relationship between $NO_x$ and VOCs and how they relate to $O_3$ production is complex. In many cases, this relationship is non-linear. Ozone isopleths, oftentimes generated using computer models or with environmental chambers, have been used to evaluate the complex relationship between $NO_x$, VOC and $O_3$ formation but these methods can be expensive and are oftentimes data-limited. If VOC concentrations are high and $NO_x$ concentrations are low, $O_3$ formation is commonly referred to as being $NO_x$-limited. In contrast, if $NO_x$ concentrations are high and VOC concentrations low, $O_3$ formation is commonly referred to as being VOC-limited. These simple relationships do break down when reactions between $NO_x$ and $O_3$ take place. An accurate assessment of what regime a region is in allows for the development of correct emission reduction strategies. For example, if a region in Utah was found to be $NO_x$-saturated [71], a reduction in VOCs would lead to a reduction in total ozone produced in the region. However, when these relationships become non-linear an ozone isopleth becomes a useful tool in studying the $NO_x$: VOC ratios to inform policies for ozone mitigation. An ozone isopleth or study of the $NO_x$: VOC ratios in the region would indicate if the relationship between $NO_x$ and VOCs is non-linear or if the region is in a transitional state between the two regimes. The confidence of the isopleths is dependent on the amount of input data. As such, these can become intensive studies. However, resources permitting, these can become powerful studies for nonattainment regions such as Salt Lake and Utah Valleys where previous studies are inconclusive.

Due to the possible intensive nature of producing enough input data for ozone isopleths, other methods for studying ozone production have been developed. Recent work by Liu et al. assessing the ratios of formaldehyde and glyoxal to $NO_2$ has shown to be successful at predicting the $O_3$-$NO_x$-VOC sensitivity [88]. Making high-time resolution measurements of formaldehyde, glyoxal and $NO_x$ allows for the determination of whether a region is in the $NO_x$ or VOC-limited regime as it relates to $O_3$ formation. The ratio of HCHO to $NO_2$ (FNR) is a good indicator of the $O_3$-$NO_x$-VOC sensitivity. [89–95] Different studies have used different FNR thresholds for determining $O_3$ formation. Recent studies have found that a FNR ratio of between 3.2 and 4.1 in the United States represents a $NO_x$-limited regime [96]. Glyoxal is commonly found in the atmosphere [88]. The main difference between HCHO and glyoxal is that HCHO concentrations are affected by primary emissions while glyoxal is mainly derived from photochemical oxidation of VOCs. Liu et al. showed that a glyoxal to $NO_2$ ratio (GNR) of <0.009 represents a VOC-limited regime and a GNR > 0.024 represents a $NO_x$-limited regime [88]. Additionally, recent published work by Kaiser et al., shows that measurement of the ratio of glyoxal to formaldehyde can provide information about hydrocarbon precursor speciation [97]. They showed that the ratio of glyoxal to formaldehyde can be a useful diagnostic of biogenic VOC emission and that the ratio of glyoxal to formaldehyde can be used to identify the speciation of VOCs leading to secondary pollutant formation. While extensive field campaigns are needed, using glyoxal and formaldehyde as surrogates to speciated VOC measurements would be far cheaper measuring speciated VOCs. The EPA only measures speciated VOCs at a handful of Photochemical Assessment Monitoring Stations (PAMS) sites which measure speciated VOCs. As data for building ozone isopleths is sparse and campaigns for $NO_x$ and speciated VOCs are expensive, measuring glyoxal and formaldehyde is a cheaper, instrumentally easier, and faster surrogate for campaigns.

While Utah has been a hub of air quality research, monitoring, and modeling, it still lacks some areas that could be improved in future research endeavors. Future research should seek to address ozone and $PM_{2.5}$ nonattainment along the Wasatch front where possible. The authors suggest strategies of source apportionment to address $PM_{2.5}$ nonattainment and VOC: $NO_x$ studies to address ozone nonattainment in Utah. These strategies would also force other pollutants mentioned in Table 15 to be studied in the process. Where possible, resources may be allocated to lesser-studied yet highly populated regions of the state such as the Utah Valley and Southern Utah regions.

**Author Contributions:** All authors contributed to the writing of the manuscript. The first draft of the manuscript was written by C.E.F. and all authors provided edits and comments on previous versions of the manuscript. All authors have read and agreed to the published version of the manuscript.

**Funding:** This work was supported by the National Science Foundation, grant # 2114655.

**Data Availability Statement:** All supporting data has been cited in the article and the data can be found at the respective locations referenced.

**Conflicts of Interest:** The authors declare no conflict of interest.